\documentclass[12pt]{iopart}

\usepackage[english]{babel}
\usepackage[utf8]{inputenc}

\usepackage{iopams}
\usepackage{mathdots,bbm,amsopn}
\newcommand{\eqref}[1]{\eref{#1}}

\usepackage{tabstackengine}
\setstackEOL{\cr}

\usepackage[hyphens]{url}
\usepackage[pdftex,bookmarks,colorlinks]{hyperref}

\usepackage[square,numbers,comma,sort&compress]{natbib}

\usepackage[inline,shortlabels]{enumitem}
\usepackage{graphicx}
\usepackage{graphbox,adjustbox}

\newcommand{\dd}[0]{\rmd}

\newcommand{\ii}[0]{\rmi}

\newcommand{\reals}{\mathbb R}
\newcommand{\complex}{\mathbb C}

\newcommand{\bra}[1]{\langle #1 |}
\newcommand{\ket}[1]{| #1 \rangle}
\newcommand{\ketbra}[2]{| #1 \rangle\!\langle #2 |}

\newcommand{\expval}[1]{\langle#1\rangle}

\newcommand{\abs}[1]{\vert#1\vert}

\newcommand{\comm}[2]{[ #1, #2 ]}

\newcommand{\acomm}[2]{\{ #1, #2 \}}

\newcommand{\quotient}{\mathcal P}
\newcommand{\rgt}[1]{\overrightarrow{#1\hspace{0.3em}}}
\newcommand{\lft}[1]{\overleftarrow{\hspace{0.3em}#1}}

\begin{document}

\title{Thermodynamic uncertainty relations for coherently driven open quantum systems}

\author{Paul Menczel$^{1, 2}$, Eetu Loisa$^{1, 3}$, Kay Brandner$^{4, 5}$ and Christian Flindt$^1$}

\address{$^1$ Department of Applied Physics, Aalto University, 00076 Aalto, Finland}
\address{$^2$ Theoretical Quantum Physics Laboratory, RIKEN Cluster for Pioneering Research, Wako-shi, Saitama 351-0198, Japan}
\address{$^3$ Department of Applied Mathematics and Theoretical Physics, University of Cambridge, Cambridge CB3 0WA, United Kingdom}
\address{$^4$ School of Physics and Astronomy, University of Nottingham, Nottingham NG7 2RD, United Kingdom}
\address{$^5$ Centre for the Mathematics and Theoretical Physics of Quantum Non-Equilibrium Systems, University of Nottingham, Nottingham NG7 2RD, United Kingdom}
\ead{paul@menczel.net}

\begin{abstract}
In classical Markov jump processes, current fluctuations can only be reduced at the cost of increased dissipation.
To explore how quantum effects influence this trade-off, we analyze the uncertainty of steady-state currents in Markovian open quantum systems.
We first consider three instructive examples and then systematically minimize the product of uncertainty and entropy production for small open quantum systems.
As our main result, we find that the thermodynamic cost of reducing fluctuations can be lowered below the classical bound by coherence.
We conjecture that this cost can be made arbitrarily small in quantum systems with sufficiently many degrees of freedom.
Our results thereby provide a general guideline for the design of thermal machines in the quantum regime that operate with high thermodynamic precision, meaning low dissipation and small fluctuations around average values.
\end{abstract}

\section{Introduction}

As the size of a thermodynamic system is reduced, the influence of thermal fluctuations becomes increasingly important.
When building nano-scale machines -- be it computer chips~\cite{ManipatruniNaturePhys2018}, molecular motors~\cite{KolomeiskyAnnuRevPhysChem2007} or quantum devices~\cite{DowlingPhilosTransRoyalSocA2003, LaddNature2010, PekolaNaturePhys2015} -- it is necessary to control and minimize these fluctuations, for example, to sustain the precision of clock signals \cite{BaratoPhysRevX2016, ErkerPhysRevX2017, MitchisonContempPhys2019} or the constancy of cooling elements \cite{PietzonkaPhysRevLett2018}.
At the same time, it is important to mitigate the dissipation in such machines to reduce the waste heat and increase their efficiency.

It was recently realized, however, that these two objectives are mutually exclusive in classical steady-state machines:
	fluctuations can only be minimized at the cost of increased dissipation and vice versa~\cite{BaratoPhysRevLett2015, HorowitzNaturePhys2020}.
This trade-off is quantified by the thermodynamic uncertainty relation, which implies that the uncertainty product
\begin{equation} \label{eq:product}
	\quotient \equiv \frac{\sigma D}{J^2}
\end{equation}
is bounded from below as
\begin{equation} \label{eq:tur}
	\quotient \geq 2 .
\end{equation}
Hence, the Fano factor $D / \abs J$, which measures the strength of the fluctuations, and the normalized dissipation rate $\sigma / \abs J$ cannot become small at the same time.
Here, $\sigma$ denotes the total entropy production rate in a steady-state thermodynamic process, $J$ the heat current into a thermal reservoir and $D$ the long-time fluctuations of the heat current.
More precisely, $J$ and $D$ are the time and ensemble averages $J \equiv \lim_{t \to \infty}\, \expval{\Delta E} / t$ and $D = \lim_{t \to \infty} \expval{ ( \Delta E - J t )^2 } / t$, where $\Delta E$ is the energy exchanged with the reservoir during a time span of length $t$.
Note that we set Boltzmann's constant to one throughout this work.

The uncertainty relation~\eqref{eq:tur} holds for all thermodynamic processes that can be described in terms of classical, time-homogeneous master equations with transition rates satisfying a local detailed balance condition \cite{BaratoPhysRevLett2015, GingrichPhysRevLett2016, GingrichJPhysA2017, SeifertPhysicaA2018, DechantArXiv180408250Cond-Matstat-Mech2019}.
As several recent works have shown, similar trade-off relations even apply in more general settings involving, for example, feedback control \cite{PottsPhysRevE2019, VuJPhysA2020}, ballistic transport \cite{BrandnerPhysRevLett2018}, periodic driving \cite{ShiraishiPhysRevLett2016, BaratoNewJPhys2018, HasegawaPhysRevLett2019, KoyukPhysRevLett2019, KoyukPhysRevLett2020, PotaninaPhysRevX2021} or finite-time observations \cite{PietzonkaPhysRevE2017, HorowitzPhysRevE2017, LiuPhysRevLett2020, FalascoNewJPhys2020}.
To compensate for their broader scope, these generalized uncertainty relations are either weaker than the original bound \eqref{eq:tur} or involve quantities other than $\sigma$, $J$ and $D$, see Refs.~\cite{SeifertAnnuRevCondensMatterPhys2019, HorowitzNaturePhys2020} for recent reviews.

In this paper, we focus on thermodynamic processes that involve Markovian open quantum systems described in terms of Lindblad master equations \cite{Breuer2002, Binder2018}.
Even though several previous works have investigated generalized thermodynamic uncertainty relations in this setting \cite{CarolloPhysRevLett2019, TimpanaroPhysRevLett2019, MenczelPhysRevResearch2020, GuarnieriPhysRevResearch2019, HasegawaPhysRevLett2020, HasegawaPhysRevLett2021, Rignon-BretPhysRevE2021, MillerPhysRevLett2021, TajimaArXiv200413412Quant-Ph2020}, the behavior of the uncertainty product~\eqref{eq:product} has not yet been systematically explored.
Our goal is thus to determine whether a trade-off between fluctuations and dissipation still holds for these systems and, if so, how much it can be alleviated by quantum effects compared to the original uncertainty relation.

We begin our analysis in Sec.~\ref{sec:examples} with three instructive setups; a spin oscillating in a magnetic field, a cavity driven by a laser field, and a three-level maser.
Here, we find that the uncertainty product indeed can be reduced below the classical bound of $2$, but it cannot be made arbitrarily small.
In Sec.~\ref{sec:general}, we carry out a systematic investigation of the lower bound of the uncertainty product as a function of the Hilbert space dimension of the open quantum system.
Specifically, we find for dimensions $2$, $3$ and $4$ that the uncertainty product is bounded from below by $1.25$, $0.47$ and $0.26$, respectively.
Finally, in Sec.~\ref{sec:conclusio}, we conclude by summarizing our results and discussing perspectives for future research.
\newpage

\section{Physical Examples} \label{sec:examples}

\subsection{Oscillating Spin} \label{subsec:spin}

We first consider the evolution of a spin-$\frac 1 2$ particle that is placed in a rotating magnetic field, which causes coherent oscillations of the spin.
Assuming that the magnetic field vector rotates with constant magnitude, the oscillations can be described by the time-dependent Hamiltonian \cite{LeBellac2006}
\begin{equation}
	H_t = \frac{\hbar\omega_0}{2} \sigma_z + \frac{\hbar\omega_1}{2} \bigl( \sigma_x \cos[\nu t] + \sigma_y \sin[\nu t] \bigr) ,
\end{equation}
where $\sigma_{x,y,z}$ are the usual Pauli matrices and $\nu$ is the angular frequency of the rotation.
The energies $\hbar\omega_0$ and $\hbar\omega_1$ are proportional to the field strengths perpendicular and parallel to the plane of rotation, respectively, times the gyromagnetic factor of the spin.
The eigenenergies $\pm\frac{\hbar\omega}{2}$ of this Hamiltonian with $\omega \equiv \sqrt{\omega_0^2 + \omega_1^2}$ are constant in time.

To obtain a realistic description of the spin-$\frac 12$ particle, we should take the thermalizing effect of its environment into account.
To provide a quanti\-tative description, we assume that both the internal relaxation of the environment and the unperturbed evolution of the spin are fast compared to the driving and coupling time scales $\nu^{-1}$ and $\hbar V^{-1}$, where $V$ is the typical system-environment interaction energy.
The time evolution of the spin state $\rho_t$ then obeys the Lindblad master equation \cite{AlickiJPhysA1979, AlbashNewJPhys2012, BrandnerPhysRevLett2020}
\begin{eqnarray}
	\partial_t \rho_t = \frac{1}{\ii\hbar} \comm{H_t}{\rho_t} &+ \gamma (n + 1)\, \bigl( L^-_t \rho_t L^+_t - \acomm{L^+_t L^-_t}{\rho_t} / 2 \bigr) \nonumber \\
		&+ \gamma n\, \bigl( L^+_t \rho_t L^-_t - \acomm{L^-_t L^+_t}{\rho_t} / 2 \bigr) . \label{eq:2ls_lindblad}
\end{eqnarray}
Here, $\acomm A B \equiv AB + BA$ denotes the anti-commutator, $\gamma$ is the characteristic thermalization rate of the spin, and the Bose-Einstein factor $n \equiv (\e^{\hbar\omega / T} - 1)^{-1}$ accounts for the finite environment temperature $T$.
The Lindblad operators $L^\pm_t \equiv \ket{\pm}_t \bra{\mp}_t$ are jump operators between the instantaneous eigenstates $\ket{\pm}_t$ of the Hamiltonian; the Lindblad equation thus satisfies the quantum detailed-balance condition \cite{GoriniRepMathPhys1978, BrandnerPhysRevE2016}.
The time-dependence of the Hamiltonian and the Lindblad operators can be removed by transforming the system to a rotating frame.
The unitary transformation
\begin{equation}
	U_t \equiv \exp\Bigl[ -\frac{\ii}{\hbar} X t \Bigr] \exp\Bigl[ \frac{\ii \sigma_y}{2} \arctan \frac{\omega_1}{\omega_0} \Bigr]
	\; \textup{ with } \;
	X \equiv \frac{\hbar\nu}{2} \Bigl( \frac{\omega_1}{\omega} \sigma_x - \frac{\omega_0}{\omega} \sigma_z \Bigr)
\end{equation}
diagonalizes the Hamiltonian, which can thus be written in the form $H_t = U_t^\dagger \bar H U_t$ with $\bar H \equiv \frac{\hbar\omega}{2} \sigma_z$ being diagonal.
The state operator in the rotating frame, $\bar\rho_t \equiv U_t \rho_t U_t^\dagger$, satisfies the time-homogeneous Lindblad equation
\begin{eqnarray}
	\partial_t \bar\rho_t = \frac{1}{\ii\hbar} \comm{\bar H + X}{\bar\rho_t} &+ \gamma (n + 1)\, \bigl( \sigma_- \bar\rho_t \sigma_+ - \acomm{\sigma_+ \sigma_-}{\bar\rho_t} / 2 \bigr) \nonumber \\
		&+ \gamma n\, \bigl( \sigma_+ \bar\rho_t \sigma_- - \acomm{\sigma_- \sigma_+}{\bar\rho_t} / 2 \bigr) \label{eq:2ls_lindblad2}
\end{eqnarray}
with $\sigma_\pm \equiv (\sigma_x \pm \ii \sigma_y) / 2$.
In this frame, the spin settles into a non-equilibrium steady state $\bar\rho_\infty$ at long times \cite{SpohnLettMathPhys1977, MenczelJPhysA2019}.

The rotating spin can be regarded as an elementary  thermal machine.
It converts the energy provided by the magnetic field into heat which is dissipated into the environment.
At long times, this heat current is given by \cite{Binder2018}
\begin{eqnarray}
	J &\equiv \hbar\omega\gamma\, \Bigl( (n + 1) \tr\bigl[ \sigma_- \bar\rho_\infty \sigma_+ \bigr] - n \tr\bigl[ \sigma_+ \bar\rho_\infty \sigma_- \bigr] \Bigr) \nonumber \\
		&= \frac{\hbar\omega\gamma\Omega^2}{\gamma^2(2n + 1)^2 + 2(\Omega^2 + 2\delta^2)} . \label{eq:tls_current}
\end{eqnarray}
In the second line, we plugged in the stationary state $\bar\rho_\infty$ and introduced the normalized driving strength $\Omega \equiv \nu\omega_1 / \omega$ and the detuning $\delta \equiv (\omega - \nu\omega_0 / \omega)$, which controls the amplitude of the spin oscillations.\footnote{
	If the influence of the environment is neglected and the initial spin state is an eigenstate of the Hamiltonian, the oscillation amplitude is maximal if and only if the detuning is zero.}
In accordance with the second law of thermodynamics, the heat current \eqref{eq:tls_current} is non-negative.
We now investigate its thermodynamic precision by determining how small the uncertainty product in Eq.~\eqref{eq:product} can become.
Note that in classical Markov processes, the validity of the thermodynamic uncertainty relation~\eqref{eq:tur} mathematically relies on two conditions \cite{BaratoPhysRevLett2015, BaratoPhysRevX2016, GingrichPhysRevLett2016}.
First, the transition rates of the Markov process are required to satisfy a local detailed balance condition.
Second, time-dependent driving protocols are not allowed, guaranteeing that the system operates in a steady state.
Since the model described by Eq.~\eqref{eq:2ls_lindblad2} satisfies detailed balance and possesses an effective steady state despite the time-dependent driving, it is a natural candidate for exploring the applicability of thermodynamic uncertainty relations in quantum systems.
Without the time-dependent driving, this model would be equivalent to a simple two-state Markov process \cite{Breuer2002} and thus completely classical.

To determine the uncertainty product, we use the framework of full counting statistics and introduce a counting field $s$.
The Lindblad equation becomes \cite{LevitovJMathPhys1996, BagretsPhysRevB2003, FlindtEPL2004}
\begin{eqnarray}
	\partial_t \bar\rho_t = \mathsf L[s] \bar\rho_t \equiv \frac{1}{\ii\hbar} \comm{\bar H + X}{\rho_t} &+ \gamma (n + 1)\, \bigl( \e^{s}\, \sigma_- \bar\rho_t \sigma_+ - \acomm{\sigma_+ \sigma_-}{\bar\rho_t} / 2 \bigr) \nonumber \\
		&+ \gamma n\, \bigl( \e^{-s}\, \sigma_+ \bar\rho_t \sigma_- - \acomm{\sigma_- \sigma_+}{\bar\rho_t} / 2 \bigr) , \label{eq:lindblad_cf}
\end{eqnarray}
where we have defined the Liouvillian $\mathsf L[s]$, which is a superoperator that acts on the space of Hermitian $2 \times 2$ matrices.
The fluctuations of the heat current can be found using the expression \cite{BrudererNewJPhys2014}
\begin{equation} \label{eq:char_poly}
	D = \frac{-1}{\partial_\lambda P[s,\lambda]} \Bigl( (\hbar\omega)^2\, \partial_s^2 + 2\hbar\omega\, J\, \partial_\lambda \partial_s + J^2\, \partial_\lambda^2 \Bigr) P[s,\lambda] \Bigr|_{s,\lambda=0} \;,
\end{equation}
where $P[s,\lambda] \equiv \det[\mathsf L[s] - \lambda \mathbbm 1]$ is the characteristic polynomial of $\mathsf L[s]$.
Since the entropy of the spin itself remains constant in the effective steady state, the total entropy production at long times stems only from the thermal environment, where entropy is generated at the rate $\sigma = J / T$.
We thus obtain the uncertainty product
\begin{eqnarray}
	\quotient &\equiv \frac{\sigma D}{J^2} = \log\Bigl[ \frac{n + 1}{n} \Bigr] \frac{D}{\hbar\omega J} \nonumber \\
		&= \log\Bigl[ \frac{n + 1}{n} \Bigr] \Bigl( 2n + 1 + \frac{2}{2n + 1} \frac{\Omega^2 [4\delta^2 - 3\gamma^2 (2n + 1)^2])}{[ \gamma^2 (2n + 1)^2 + 2(\Omega^2 + 2\delta^2) ]^2} \Bigr) , \label{eq:2ls_Q}
\end{eqnarray}
having inserted Eqs.~\eqref{eq:tls_current} and \eqref{eq:char_poly} in the second line and evaluated the derivatives.

To see if our setup can violate the classical relation $\quotient \geq 2$, we minimize the uncertainty product as a function of the system parameters.
It becomes minimal for zero detuning, Bose-Einstein factor $n \approx 0.02779$ and a driving strength of $\Omega = 2^{-1/2} (2n + 1) \gamma$.
The exact value of $n$ is the positive root of the transcendental equation
\begin{equation}
	(2n + 1) (16n^2 + 16n + 1) = 2n (n + 1) (16n^2 + 16n + 7) \log\Bigl[ \frac{n + 1}{n} \Bigr] .
\end{equation}
At these parameter values, the uncertainty product reaches the value
\begin{equation} \label{eq:2ls_min}
	\quotient_{\mathrm{min}}^{\mathrm{(spin)}} \approx 1.2459 < 2 .
\end{equation}
The coherence stemming from the non-diagonal effective Hamiltonian $\bar H + X$ can thus decrease the uncertainty product below the classical limit, but it cannot become arbitrarily small:
	the spin system still satisfies the weaker thermo\-dynamic uncertainty relation $\quotient \geq \quotient_{\mathrm{min}}^{\mathrm{(spin)}}$.
The increase in precision compared to classical systems can be considered a genuine quantum effect, since
\begin{equation}
	\quotient = (2n + 1) \log\Bigl[ \frac{n + 1}{n} \Bigr] + \mathcal O(\Omega^2) > 2
\end{equation}
in the semi-classical limit, where $\Omega$ is small.

\begin{figure}
	\centering
	\includegraphics{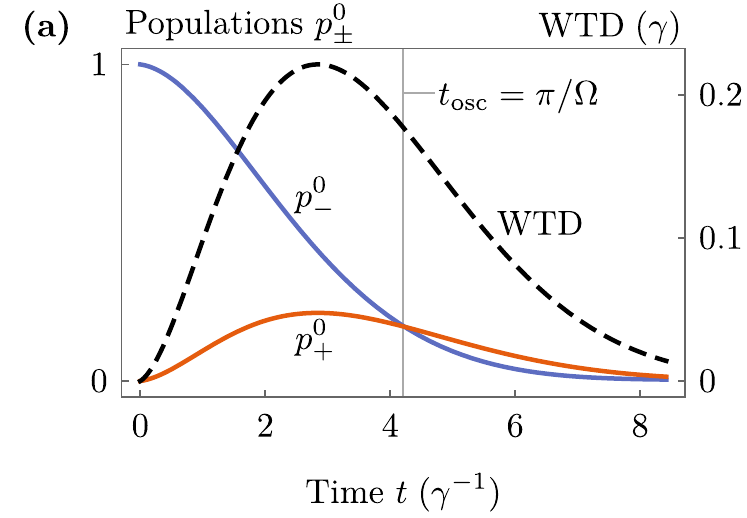}
	\;
	\includegraphics{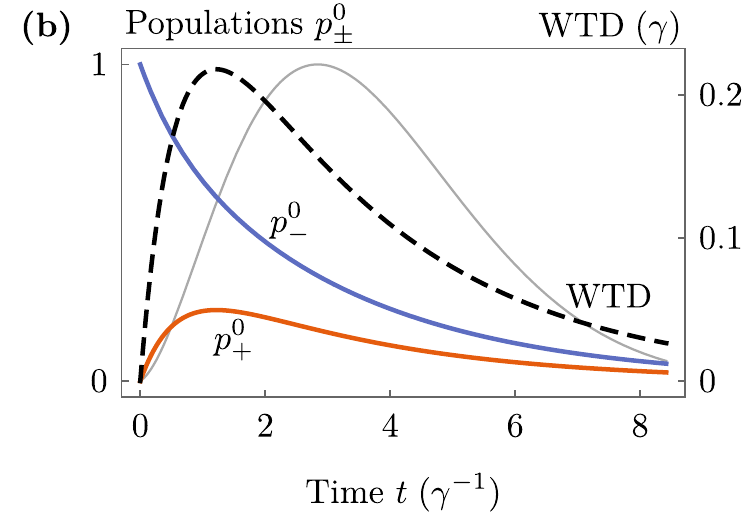}
	\caption[.]{\begin{enumerate*}[\textbf{(\alph*)}]
			\item Time evolution of the spin after an emission event in the optimal setup.
				The thick solid curves show the conditional populations $p^0_\pm$ and are marked on the left vertical axis.
				The dashed curve is the waiting time distribution (WTD) and is marked on the right vertical axis.
				The time $t_{\mathrm{osc}}$ is indicated with a thin vertical line.
			\item Time evolution of the SET after the emission of an electron into the right lead.
				The thick solid curves and the dashed curve correspond to the curves in panel~(a).
				The thin curve is the waiting time distribution from panel~(a) for the sake of comparison.
		\end{enumerate*}}
	\label{fig:2ls}
\end{figure}

We conclude this discussion with an analysis of the mechanisms behind the optimal setup with the minimal uncertainty product.
The magnetic field repeatedly drives the spin from the ground state to the excited state, from where it decays back into the ground state by emitting an energy quantum of size $\hbar\omega$ into the environment.
This process is illustrated in Fig.~\ref{fig:2ls}(a), where we assume that the spin is initially in the ground state and we show the conditional populations as well as the emission waiting time distribution (WTD) as a function of time.
The conditional populations $p^0_\pm = p^0_\pm[t]$ describe the probability for no spontaneous decay to happen until the time $t$ and for the spin to be found in the eigenstate $\ket{\pm}$ at that time.
The emission WTD is the statistical distribution of the time of the first decay event. 
These quantities can be calculated using methods of full counting statistics \cite{PlenioRevModPhys1998, CarmichaelPhysRevA1989, AlbertPhysRevLett2012, HaackPhysRevB2014}, see \ref{app:2ls_set}.
The curves show that the coherent oscillations invert the populations on the time scale $t_{\mathrm{osc}} = \pi / \Omega$, and that the waiting time distribution correspondingly is peaked around $t_{\mathrm{osc}}$.

It is instructive to compare these results with a semi-classical process such as the charge transport through a single-electron transistor (SET) coupled to two leads.
In the Coulomb-blockade regime, the SET corresponds to a classical two-level system, since only two of its charge states are accessible.
We denote their populations by $p_+$ and $p_-$, which satisfy the classical master equation \cite{Schaller2015}
\begin{equation} \label{eq:set}
	\partial_t \pmatrix{ {p_+} \cr {p_-} } = \gamma\, \pmatrix{ -(1 - f_L) - (1 - f_R) & +f_L + f_R \cr +(1 - f_L) + (1 - f_R) & -f_L - f_R } \pmatrix{ {p_+} \cr {p_-} } .
\end{equation}
The Fermi-Dirac factors $f_{L,R} = [\e^{(\hbar\omega \mp eV/2) / T_{\mathrm{SET}}} - 1]^{-1}$ of the left ($-$) and right ($+$) leads depend on the temperature $T_{\mathrm{SET}}$ and on the applied voltage bias $V$.
The rate~$\gamma$ and the energy $\hbar\omega$ set the overall time and energy scales; for illustrative purposes, we identify them with the constants $\gamma$ and $\hbar\omega$ used before.
The temperature and the applied voltage are chosen so that both the long-time current into the right lead and the mean of the waiting time distribution of  emissions into the right lead coincide with the spin-$\frac 12$ setup.
The current into the right lead then has the uncertainty product $\quotient \approx 2.93$.
Figure~\ref{fig:2ls}(b) shows the emission WTD of the SET as well as the time evolution of its conditional populations assuming no emission into the right lead, see \ref{app:2ls_set}.

A comparison of the two WTDs shows that the one of the SET is characterized by a peak at shorter times and a long tail.
It thus has a higher variance, corresponding to the higher uncertainty product.
The WTD of the optimal spin setup is narrower because the resonant coherent oscillations are able to fully invert the populations, decreasing the likelihood of very long waiting times.
However, the fluctuations and the uncertainty product in the spin setup are still larger than zero.
The remaining fluctuations stem from the stochastic nature of the emission process and from the finite environment temperature, which is responsible for random excitations of the spin.
The optimal Bose-Einstein factor $n$ can thus be understood as the result of a trade-off between low temperatures, which increase the entropy production in the environment, and high temperatures, which increase the fluctuations by making excitation events more likely.
The optimal driving strength $\Omega$ results from a similar trade-off:
	too slow coherent oscillations have a higher probability of excitation events, and too fast oscillations increase the chance for the device to miss a cycle and return to the ground state without the emission of an energy quantum into the reservoir.

\subsection{Driven Cavity} \label{subsec:cavity}

Next, we consider a light mode in an optical cavity which is driven by a classical laser field.
Formally, this setup is rather similar to the oscillating spin model discussed above.
However, we replace the spin ladder operators $\sigma_\pm$ in Eq.~\eqref{eq:2ls_lindblad2} with the bosonic ladder operators $a$ and $a^\dagger$ with $\comm{a}{a^\dagger} = 1$ to obtain the master equation \cite{Carmichael2008, Hofer2016}
\begin{eqnarray} \label{eq:ho:lindblad}
	\partial_t \bar\rho_t = -\ii\, \comm{\delta\, a^\dagger a + \Omega\, (a + a^\dagger)}{\bar\rho_t} &+ \gamma (n + 1)\, \bigl( a \bar\rho_t a^\dagger - \acomm{a^\dagger a}{\bar\rho_t} / 2 \bigr) \nonumber \\
		&+ \gamma n\, \bigl( a^\dagger \bar\rho_t a - \acomm{a a^\dagger}{\bar\rho_t} / 2 \bigr) .
\end{eqnarray}
Here $\bar\rho_t$ is the state of the cavity mode in a rotating frame and the rates $\Omega$, $\delta$ and $\gamma$ parameterize the driving strength, the detuning, and the dissipation rate, respectively.
The Bose-Einstein factor $n \equiv (\e^{\hbar\omega / T} - 1)^{-1}$ depends on the mode frequency $\omega$ and on the temperature $T$ of the environment.
In the absence of the external driving field, it would be equal to the mean thermal occupation of the mode.

To find the stationary state, we introduce the displaced ladder operators
\begin{equation} \label{eq:ho:b}
	b \equiv a + \lambda
	\; \textup{ and }\;
	b^\dagger = a^\dagger + \lambda^\ast
	\quad \textup{with} \quad
	\lambda \equiv \frac{2\ii\, \Omega}{\gamma + 2\ii\, \delta} \in \complex ,
\end{equation}
satisfying $\comm{b}{b^\dagger} = 1$.
Using this definition, Eq.~\eqref{eq:ho:lindblad} takes the familiar form of the master equation for a non-driven damped harmonic oscillator,
\begin{eqnarray}
	\partial_t \bar\rho_t = -\ii\delta\, \comm{b^\dagger b}{\bar\rho_t} &+ \gamma (n + 1)\, \bigl( b \bar\rho_t b^\dagger - \acomm{b^\dagger b}{\bar\rho_t} / 2 \bigr) \nonumber \\
		&+ \gamma n\, \bigl( b^\dagger \bar\rho_t b - \acomm{b b^\dagger}{\bar\rho_t} / 2 \bigr) .
\end{eqnarray}
The stationary state is therefore given by the thermal state in terms of the displaced ladder operators,
\begin{equation} \label{eq:ho:stat_state}
	\bar\rho_\infty = Z^{-1} \exp\bigl[ -(\hbar\omega / T)\, b^\dagger b \bigr] ,
\end{equation}
where $Z \equiv \tr\exp[ -(\hbar\omega / T)\, b^\dagger b ]$ is the partition function.
We use it to obtain the long-time heat current from the cavity into the environment as
\begin{equation} \label{eq:ho:current}
	J \equiv \hbar\omega\gamma\, \Bigl( (n + 1) \tr\bigl[ a \bar\rho_\infty a^\dagger \bigr] - n \tr\bigl[ a^\dagger \bar\rho_\infty a \bigr] \Bigr) \nonumber = \hbar\omega\gamma\, \frac{4\Omega^2}{\gamma^2 + 4\delta^2} .
\end{equation}

In \ref{app:ho}, we derive the corresponding fluctuations $D = \hbar\omega J\, (2n + 1)$.
The uncertainty product is thus given by
\begin{equation} \label{eq:ho:Q}
	\quotient = \log\Bigl[ \frac{n + 1}{n} \Bigr] \frac{D}{\hbar\omega J} = (2n + 1) \log\Bigl[ \frac{n + 1}{n} \Bigr] > 2 .
\end{equation}
Since this expression is strictly larger than two, the setup obeys the classical thermo\-dynamic uncertainty relation.
This result reveals a surprisingly big difference between the oscillating spin and the driven cavity, despite their formal similarity.
In fact, the long-time behavior of the cavity is fully classical:
	as we show in \ref{app:ho}, the scaled long-time cumulants $c_k$ of the heat current (with $c_1 = J$ and $c_2 = D$) have the form
	\begin{equation} \label{eq:ho:cumulants}
		c_k = \gamma\abs{\lambda}^2 (\hbar\omega)^k\, (1 + n + (-1)^k n) .
	\end{equation}
These cumulants agree with those of a classical, one-dimensional discrete random walk.
In other words, at long times, the emission and absorption processes can be understood as independent Poisson processes with characteristic rates $\gamma\abs{\lambda}^2 (n{+}1)$ and $\gamma\abs{\lambda}^2 n$, respectively.
Hence, the uncertainty product \eqref{eq:ho:Q} coincides with that of a classical biased random walk~\cite{BaratoPhysRevLett2015}.

\subsection{Three-Level Maser} \label{subsec:maser}

In Sec.~\ref{subsec:spin}, we demonstrated that it is  possible for quantum devices to operate with high thermodynamic precision, violating the classical thermodynamic uncertainty relation.
However, the thermal machine considered there does not perform a useful task: it only converts the energy supplied by the external drive into waste heat.
In this section, we consider the three-level maser as an example of a system which violates the classical uncertainty relation and acts as a useful quantum heat engine by converting thermal energy into work.

The fact that a three-level maser can be understood as a quantum heat engine was realized in 1959 by Scovil and Schulz-DuBois \cite{ScovilPhysRevLett1959}.
The working substance of the engine is a gas of three-level atoms, which is coupled to a hot and a cold thermal reservoir.
We denote the three states by $\ket n$ ($n \in \{1, 2, 3\}$) and the corresponding energies by $E_n$ with $E_1 < E_2 < E_3$.
The coupling to the reservoirs can be designed so that the hot reservoir only induces transitions between states $\ket 1$ and $\ket 3$, and the cold reservoir only between $\ket 2$ and $\ket 3$ \cite{ZouPhysRevLett2017, KlatzowPhysRevLett2019}. 
This configuration creates population inversion between states $\ket 1$ and $\ket 2$, which can be exploited to extract work from the system.
Specifically, the work is extracted in the form of photons emitted into a light mode, which drives coherent oscillations between states $\ket 1$ and $\ket 2$.
The time evolution of the three-level atoms is described by the Lindblad master equation \cite{KlatzowPhysRevLett2019}
\begin{eqnarray}
	\partial_t \bar\rho_t = &{-\ii}\, \comm{\nu \sigma_{33} + \Omega \sigma_{12} + \Omega \sigma_{21}}{\bar\rho_t} \nonumber \\
		&+ \gamma (n + 1)\, \bigl( \sigma_{32} \bar\rho_t \sigma_{23} - \acomm{\sigma_{33}}{\bar\rho_t} / 2 \bigr) + \gamma n\, \bigl( \sigma_{23} \bar\rho_t \sigma_{32} - \acomm{\sigma_{22}}{\bar\rho_t} / 2 \bigr) \nonumber \\
		&+ \Gamma\, \bigl( \sigma_{31} \bar\rho_t \sigma_{13} - \acomm{\sigma_{33}}{\bar\rho_t} / 2 \bigr) + \Gamma\, \bigl( \sigma_{13} \bar\rho_t \sigma_{31} - \acomm{\sigma_{11}}{\bar\rho_t} / 2 \bigr) . \label{eq:maser:lindblad}
\end{eqnarray}
Here, $\bar\rho_t$ is the system state in a rotating frame, the frequency $\Omega$ corresponds to the driving strength, and we have defined $\sigma_{ij} \equiv \ketbra j i$ and $\hbar\nu \equiv E_3 - (E_1 + E_2) / 2$.
For the sake of simplicity, we assume that the driving is resonant and that the temperature of the hot reservoir is large compared to the energy difference $E_3 - E_1$.
The excitation and decay rates associated with the hot reservoir are hence the same, and we denote them by $\Gamma$.
Finally, $\gamma$ is the characteristic interaction rate of the cold reservoir and $n \equiv (\e^{(E_3 - E_2) / T_c} - 1)^{-1}$ parameterizes its temperature $T_c$.

The stationary state $\bar\rho_\infty$ of the master equation \eqref{eq:maser:lindblad} can be determined easily, and we then obtain the heat current into the cold reservoir as
\begin{eqnarray}
	J_c &\equiv \gamma(E_3 - E_2)\, \Bigl( (n + 1) \tr\bigl[ \sigma_{32} \bar\rho_\infty \sigma_{23} \bigr] - n \tr\bigl[ \sigma_{23} \bar\rho_\infty \sigma_{32} \bigr] \Bigr) \nonumber \\
		&= \frac{4\gamma\Gamma (E_3 - E_2)\, \Omega^2}{\gamma\Gamma (\gamma n + \Gamma) (3n + 1) + 4\Omega^2 (3\Gamma + \gamma (3n + 2))} . \label{eq:maser:current}
\end{eqnarray}
An analogous calculation yields the heat current running from the hot reservoir into the system, $J_h = \frac{E_3 - E_1}{E_3 - E_2} J_c$.
Using the first law of thermodynamics, the output power is
\begin{equation}
	P \equiv J_h - J_c = \frac{E_2 - E_1}{E_3 - E_2} J_c .
\end{equation}
Since this expression is non-negative, the machine acts as a quantum heat engine for all values of the system parameters.
Its thermodynamic efficiency
\begin{equation}
	\eta \equiv \frac{P}{J_h} = \frac{E_2 - E_1}{E_3 - E_1}
\end{equation}
takes values between zero and the Carnot efficiency, which is one here, since the temperature of the hot reservoir is effectively infinite.

Due to the structure of the Lindblad equation \eqref{eq:maser:lindblad}, the number of excitations induced by the hot reservoir is equal to the number of decay events induced by the cold reservoir at long times, and vice versa \cite{MitchisonContempPhys2019}.
The thermodynamic precision of the heat currents $J_c$ and $J_h$ is therefore identical.
To find the corresponding uncertainty product $\quotient \equiv \sigma D_c / J_c^2\; (= \sigma D_h / J_h^2)$, we first note that the total entropy production rate in the stationary state is $\sigma = J_c / T_c$, because the entropy production in the hot reservoir vanishes in the limit of infinite hot temperature.
After introducing a counting field, the fluctuations $D_c$ can be determined using the formula \eqref{eq:char_poly} in analogy to the first example.
The resulting expression $\quotient$ is too complicated to be reproduced here, but it can be minimized using numerical methods.
We find that the minimal uncertainty product is
\begin{equation} \label{eq:maser_bound}
	\quotient_{\mathrm{min}}^{\mathrm{(maser)}} \approx 1.6180 ,
\end{equation}
which is below the classical bound.
The minimum is achieved for $\Omega^2 \approx 0.1277\, \Gamma^2$, $\gamma \approx 5.855\, \Gamma$ and $n \approx 0.02204$ (with $\Gamma$ and $\nu$ being free parameters).

In Ref.~\cite{PietzonkaPhysRevLett2018}, it was demonstrated that the classical thermodynamic uncertainty relation implies a trade-off relation between the power and the efficiency of classical steady-state heat engines.
Starting from our result $\sigma D_c / J_c^2 \geq \quotient_{\mathrm{min}}^{\mathrm{(maser)}}$, we can repeat their derivation to obtain an analogous trade-off relation for the three-level maser,
\begin{equation} \label{eq:power_efficiency}
	\quotient_{\mathrm{min}}^{\mathrm{(maser)}} P \leq \frac{D_P}{T_c} \frac{1 - \eta}{\eta} ,
\end{equation}
where $D_P \equiv \bigl( \frac{E_2 - E_1}{E_3 - E_2} \bigr)^2 D_c$ denotes the fluctuations of the output power.
This inequality shows that for a fixed output power, the efficiency can only be increased at the cost of increasing the power fluctuations.
Compared with a classical steady-state heat engine, this cost is reduced by a factor of $\quotient_{\mathrm{min}}^{\mathrm{(maser)}} / 2 \approx 0.81$.
For both the three-level maser and classical engines, these trade-off relations imply that the Carnot efficiency can only be reached at finite power in the limit of diverging power fluctuations.
Note that even if finite temperatures of the hot reservoir are admitted, the uncertainty product cannot be lowered further than the bound \eqref{eq:maser_bound}; the trade-off \eqref{eq:power_efficiency} therefore still applies in this case.

\section{Minimizing the Uncertainty Product} \label{sec:general}

\subsection{General Setup} \label{subsec:general}

We now consider a general $N$-level quantum system and assume that the time evolution of the system state $\rho_t$ is given by a Lindblad master equation of the form
\begin{eqnarray}
	\partial_t \rho_t = \frac{1}{\ii\hbar} \comm{H}{\rho_t} &+ \sum\nolimits_\mu \gamma_\mu (n_\mu + 1)\, \bigl( L_\mu \rho_t L_\mu^\dagger - \acomm{L_\mu^\dagger L_\mu}{\rho_t} / 2 \bigr) \nonumber \\
		&+ \sum\nolimits_\mu \gamma_\mu n_\mu\, \bigl( L_\mu^\dagger \rho_t L_\mu - \acomm{L_\mu L_\mu^\dagger}{\rho_t} / 2 \bigr) . \label{eq:general:lindblad}
\end{eqnarray}
Here, the index $\mu$ enumerates the dissipation channels with their respective rates $\gamma_\mu > 0$, Lindblad operators $L_\mu$ and Bose-Einstein factors $n_\mu$.
The effective Hamiltonian $H \equiv H_0 + X$ is the sum of a free Hamiltonian $H_0$ and a driving contribution $X$.
The Lindblad operators are lowering operators with respect to $H_0$, meaning that they satisfy
\begin{equation} \label{eq:general:db}
	\comm{H_0}{L_\mu} = -\Delta E_\mu L_\mu
\end{equation}
for some energy difference $\Delta E_\mu \geq 0$.
Hence, we do not demand that the dissipative terms adhere to the full quantum detailed-balance condition, which would require the Lindblad operators to be lowering operators with respect to the full Hamiltonian $H$.
As we have seen in Sec.~\ref{sec:examples}, the type of Lindblad equation considered here arises, for example, when a system that obeys the full detailed-balance condition is described in a rotating frame.
This type of Lindblad equation is also used for open quantum systems that consist of multiple subsystems, which are each connected to individual thermal reser\-voirs and only weakly coupled to each other \cite{RivasNewJPhys2010, ProsenPhysRevLett2011, KarevskiPhysRevLett2013, Rignon-BretPhysRevE2021}.
In the following, we investigate this type of setup without explicit reference to any particular physical implementation.

To discuss the thermodynamic behavior of the system at long times, we assume that it eventually reaches a unique stationary state $\rho_\infty$.
The particle current flowing from the open quantum system into the environment via the $\mu$-th dissipation channel is then given by
\begin{equation}
	j_\mu \equiv \gamma_\mu (n_\mu + 1) \tr[ L_\mu \rho_\infty L_\mu^\dagger ] - \gamma_\mu n_\mu \tr[ L_\mu^\dagger \rho_\infty L_\mu ] ,
\end{equation}
and it carries the heat current $J_\mu \equiv \Delta E_\mu\, j_\mu$.
This identification of heat guarantees the validity of the second law of thermodynamics, such that the total entropy production rate
\begin{equation} \label{eq:general:entropy}
	\sigma \equiv \sum\nolimits_\mu J_\mu / T_\mu = \sum\nolimits_\mu j_\mu \log\Bigl[ \frac{n_\mu + 1}{n_\mu} \Bigr] \geq 0
\end{equation}
is non-negative \cite{ChiaraNewJPhys2018}.
Here, $T_\mu$ is the temperature corresponding to the $\mu$-th dissipation channel with $n_\mu \equiv (\e^{\Delta E_\mu / T_\mu} - 1)^{-1}$.

For each dissipation channel with corresponding current $j_\mu$, we define the uncer\-tainty product $\quotient_\mu$ to be
\begin{equation} \label{eq:general:product}
	\quotient_\mu \equiv \frac{\sigma d_\mu}{j_\mu^2} .
\end{equation}
To calculate the long-time currents $j_\mu$ and the respective fluctuations $d_\mu$, we add a counting field $s_\mu$ for each dissipation channel as follows,
\begin{eqnarray}
	\partial_t \rho_t = \mathsf L[s] \rho_t \equiv \frac{1}{\ii\hbar} \comm{H}{\rho_t} &+ \sum\nolimits_\mu \gamma_\mu (n_\mu + 1)\, \bigl( \e^{s_\mu}\, L_\mu \rho_t L_\mu^\dagger - \acomm{L_\mu^\dagger L_\mu}{\rho_t} / 2 \bigr) \nonumber \\
		&+ \sum\nolimits_\mu \gamma_\mu n_\mu\, \bigl( \e^{-s_\mu}\, L_\mu^\dagger \rho_t L_\mu - \acomm{L_\mu L_\mu^\dagger}{\rho_t} / 2 \bigr) .
\end{eqnarray}
From the characteristic polynomial of $\mathsf L[s]$, $P[s,\lambda] \equiv \det[\mathsf L[s] - \lambda \mathbbm 1]$, we obtain \cite{BrudererNewJPhys2014}
\begin{eqnarray}
	j_\mu &= -\frac{\partial_{\mu} P[s,\lambda]}{\partial_\lambda P[s,\lambda]} \Biggr|_{s,\lambda=0}
		\; \textup{ and } \nonumber \\
	d_\mu &= \frac{-1}{\partial_\lambda P[s,\lambda]} \Bigl( \partial_\mu^2 + 2j_\mu\, \partial_\lambda \partial_\mu + j_\mu^2\, \partial_\lambda^2 \Bigr) P[s,\lambda] \Bigr|_{s,\lambda=0} \;. \label{eq:general:char_poly}
\end{eqnarray}
Here, the partial derivative with respect to $s_\mu$ is abbreviated as $\partial_\mu$. 

Our goal is to find the minimum uncertainty product $\quotient_{\mathrm{min}}^N$ for a fixed system dimension $N$.
Hence, we must minimize Eq.~\eqref{eq:general:product} over all dissipation channels in all possible Lindblad equations of the form \eqref{eq:general:lindblad}.
Before carrying out this minimization for the dimensions $N=2$, $3$, and $4$, we make three general remarks.
First, we observe that neither the eigenenergies of the free Hamiltonian nor its level splittings $\Delta E_\mu$ enter the definition of the uncertainty products, since we consider the factors $n_\mu$ to be free parameters.
The postulated structure $H = H_0 + X$ therefore only serves to restrict the set of acceptable Lindblad operators $L_\mu$ and has no other effect on the thermodynamic precision.
Second, we note that the uncertainty products do not change if we apply a unitary transformation $V$ to all operators in the Lindblad equation through the replacements
\begin{equation} \label{eq:general:symmetry}
	H \to V^\dagger H V
	\qquad \textup{and} \qquad
	L_\mu \to V^\dagger L_\mu V .
\end{equation}
This symmetry will be important to reduce the parameter space in our search for the minimum uncertainty product.
Finally, we remark that $\quotient_{\mathrm{min}}^N$ decreases mono\-tonically as a function of the system dimension $N$.
In order to prove this claim, we consider an $N$-dimensional quantum system which minimizes the uncertainty product at that dimension.
We thus have $\quotient_\mu = \quotient_{\mathrm{min}}^N$ for one of its dissipation channels $\mu$.
We will now construct an $(N{+}1)$-dimensional system with the same uncertainty product.
To this end, we add an additional state $\ket{N{+}1}$ with $H_0 \ket{N{+}1} = E_{N{+}1} \ket{N{+}1}$ to the system, where $E_{N{+}1}$ is chosen larger than all other eigenenergies of $H_0$.
All Lindblad operators as well as the driving operator $X$ are extended with zeroes in the additional row and column.
We also add an additional dissipation channel $\nu$ with Lindblad operator $L_\nu = \ketbra{1}{N{+}1}$.
In the limit $n_\nu \to 0$, the stationary state of the extended system lies entirely within the original $N$-dimensional subspace and is identical to the stationary state of the original system.
Hence, $\quotient_\mu = \quotient_{\mathrm{min}}^N$ still holds, which concludes our proof.

\subsection{Two-Level Systems} \label{subsec:2ls}

We begin our general analysis by considering two-dimensional open quantum systems.
To determine the configuration with the minimum uncertainty product, we first investigate which forms the dissipative terms of the Lindblad equation can take for a two-level system.
We then carry out the minimization and show that the optimal configuration is given by the oscillating spin setup that was already discussed in Sec.~\ref{subsec:spin}; the lower bound of the uncertainty product is therefore
\begin{equation}
	\quotient_{\mathrm{min}}^2 = \quotient_{\mathrm{min}}^{\mathrm{(spin)}} \approx 1.2459 .
\end{equation}

We denote the eigenstates of the free Hamiltonian $H_0$ by $\ket +$ and $\ket -$ and the respective eigenenergies by $E_+$ and $E_-$ with $E_+ > E_-$.
The Lindblad operators must satisfy condition \eqref{eq:general:db}, which has only two non-trivial solutions:
\begin{equation}
	L_{\mathrm{th}} \equiv \sigma_- \equiv \ketbra - +
	\qquad \textup{and} \qquad
	L_{\mathrm{dp}} \equiv \sigma_z \equiv \ketbra + + - \ketbra - - .
\end{equation}
The latter solution corresponds to a pure dephasing channel, which does not carry any heat current.
We thus assume that there is at least one dissipation channel with Lindblad operator $L_{\mathrm{th}}$ present and we then evaluate the associated uncertainty product.
If~$K$ denotes the number of additional dissipation channels with Lindblad operator $L_{\mathrm{th}}$ and $M$ the number of channels with Lindblad operator $L_{\mathrm{dp}}$, the Lindblad equation is
\begin{equation} \label{eq:2ls_gen:lindblad1}
	\partial_t \rho_t = \frac{1}{\ii\hbar} \comm{H}{\rho_t}
		+ \gamma_1 \mathsf D_{\mathrm{th}}[n_1]\, \rho_t
		+ \sum_{k=1}^K \gamma^{\mathrm{th}}_k \mathsf D_{\mathrm{th}}[n^{\mathrm{th}}_k]\, \rho_t
		+ \sum_{m=1}^M \gamma^{\mathrm{dp}}_m \mathsf D_{\mathrm{dp}}[n^{\mathrm{dp}}_m]\, \rho_t ,\hspace{.5em}
\end{equation}
where we have defined the dissipation superoperators
\begin{equation} \label{eq:dissipator}
	\mathsf D_i[n]\, \rho_t \equiv (n+1) \bigl( L_i \rho_t L_i^\dagger - \acomm{L_i^\dagger L_i}{\rho_t} / 2 \bigr) + n \bigl( L_i^\dagger \rho_t L_i - \acomm{L_i L_i^\dagger}{\rho_t} / 2 \bigr) .\hspace{.5em}
\end{equation}

Since our goal is to determine the minimum uncertainty product, we are free to make any modification of this equation that does not increase $\quotient_1$.
For instance, the two sums over dissipation channels can each be replaced by a single, effective channel as follows.
For the sum over $k$, we substitute the term $\gamma_2 \mathsf D_{\mathrm{th}}[n_2]\, \rho_t$ corresponding to a dissipation channel with coupling rate $\gamma_2 \equiv \sum_k \gamma^{\mathrm{th}}_k$ and Bose-Einstein coefficient $n_2 \equiv \sum_k \gamma^{\mathrm{th}}_k n^{\mathrm{th}}_k / \gamma_2$.
The sum over the pure dephasing channels is replaced with the term $\gamma_3 \mathsf D_{\mathrm{dp}}[0]\, \rho_t$ for $\gamma_3 \equiv \sum_m \gamma^{\mathrm{dp}}_m (2n^{\mathrm{dp}}_m + 1)$.
Both of these replacements do not change the right-hand side of Eq.~\eqref{eq:2ls_gen:lindblad1} and, therefore, do not change $j_1$ or $d_1$.
Due to the convexity of the function
\begin{equation}
	f(n) \equiv \Bigl( (n + 1) \tr[ L_{\mathrm{th}} \rho_t L_{\mathrm{th}}^\dagger ] - n \tr[ L_{\mathrm{th}}^\dagger \rho_t L_{\mathrm{th}} ] \Bigr) \log\Bigl[ \frac{n+1}{n} \Bigr]
\end{equation}
for $n > 0$,
	the entropy production rate of the modified system cannot be larger than that of the original system.
We have thus shown that our replacements cannot increase~$\quotient_1$.
We further note that the Hamiltonian can be brought into the form $H = \hbar\Omega\, \sigma_x + \hbar\delta\, \sigma_z$ using a symmetry transformation of the type \eqref{eq:general:symmetry}, where $V = \exp[\ii\varphi \sigma_z]$ and $\varphi$ is a suitably chosen phase.
This transformation does not modify the dissipative terms of the Lindblad equation, since it changes the Lindblad operators only by an overall phase.
In order to minimize the uncertainty product over all two-level systems, it therefore suffices to only consider Lindblad equations of the form
\begin{equation} \label{eq:2ls_gen:lindblad2}
	\partial_t \rho_t = -\ii \comm{\Omega \sigma_x + \delta \sigma_z}{\rho_t}
		+ \gamma_1 \mathsf D_{\mathrm{th}}[n_1]\, \rho_t
		+ \gamma_2 \mathsf D_{\mathrm{th}}[n_2]\, \rho_t
		+ \gamma_3 \mathsf D_{\mathrm{dp}}[0]\, \rho_t .
\end{equation}

\begin{figure}
	\centering
	\includegraphics[align=t,vshift=.75em]{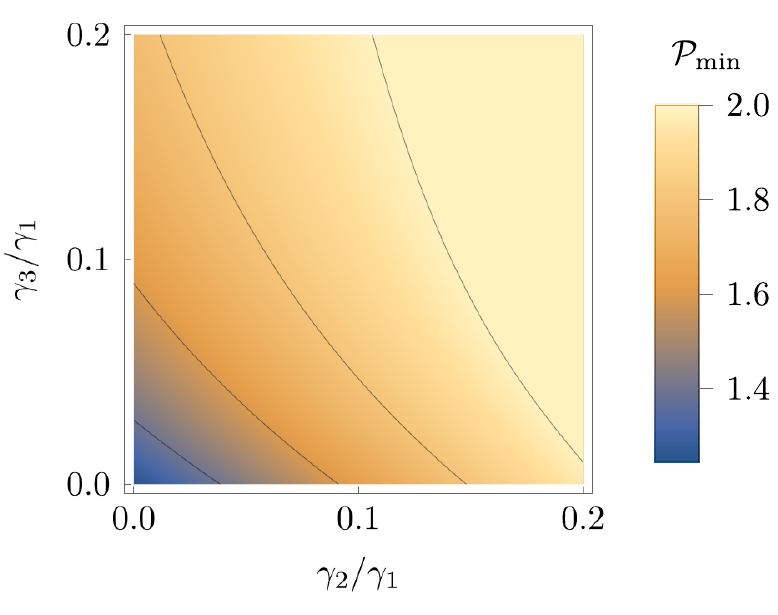}
	\caption[.]{Numerical minimization of the uncertainty product in two-level systems, based on the Lindblad equation \eqref{eq:2ls_gen:lindblad2}.
		For each fixed value of $\gamma_2 / \gamma_1$ and $\gamma_3 / \gamma_1$, the uncertainty product $\quotient_1$ was numerically minimized over all remaining dimensionless parameters, i.e., $\Omega / \gamma_1$, $\delta / \gamma_1$, $n_1$ and $n_2$.
		The plot shows the resulting value $\quotient_{\mathrm{min}}$.
		The thin contour lines indicate the values $\quotient_{\mathrm{min}} = 1.4$, $1.6$, $1.8$ and $2$ from left to right.
		Beyond the rightmost contour line, the function is constant and equal to $2$.
		The minimum $\quotient_{\mathrm{min}}^2 \approx 1.2459$ is reached for $\gamma_2$ and $\gamma_3$ going to zero.}
	\label{fig:general:2ls}
\end{figure}

This equation differs from the master equation \eqref{eq:2ls_lindblad2} in Sec.~\ref{subsec:spin} only by the addition of two extra dissipation channels.
To analyze their effect, we numerically minimize the uncertainty product $\quotient_1$ for fixed ratios $\gamma_2 / \gamma_1$ and $\gamma_3 / \gamma_1$ over all remaining parameters.
The results in Fig.~\ref{fig:general:2ls}(a) clearly show that the minimum uncertainty product is reached for $\gamma_2, \gamma_3 \to 0$.
The addition of dissipation channels thus always leads to a loss of thermodynamic precision in two-level systems, and the minimum uncertainty product is $\quotient_{\mathrm{min}}^2 = \quotient_{\mathrm{min}}^{\mathrm{(spin)}}$.
This finding concludes our investigation of two-level systems.

\subsection{Three-Level Systems}

Having determined the lower bound of the uncertainty product in two-level quantum systems, we now move on to three-level systems.
Since $\quotient_{\mathrm{min}}^N$ decreases monotonically with the system dimension, the optimal three-level system cannot be a three-level maser as discussed in Sec.~\ref{subsec:maser}.
To find the optimal system, we first analyze the possible types of Lindblad operators in three-level open quantum systems.
Let $\ket j$ for $1 \leq j \leq 3$ be the eigenstates of the free Hamiltonian and $E_j$ the corresponding eigenenergies, ordered such that $E_1 < E_2 < E_3$.
Ignoring pure dephasing channels from now on, the condition~\eqref{eq:general:db} then only permits the Lindblad operators
\begin{equation}
	L_a \equiv \ketbra 1 2, \;
	L_b \equiv \ketbra 2 3, \;
	L_c \equiv \ketbra 1 3 \; \textup{ and } \;
	L_d[\alpha] \equiv L_1 + \alpha L_2 ,
\end{equation}
where $\alpha \neq 0$ is a complex parameter.
We note that dissipation channels with Lindblad operators of the type $L_d[\alpha]$ are only permitted if the condition $E_3 - E_2 = E_2 - E_1$ is satisfied.
However, since the uncertainty product does not depend on the eigenenergies $E_j$, we are free to choose them so that the condition holds.

In our discussion of two-level systems, we found that the addition of extra dissipation channels generally increases the uncertainty products.
We therefore restrict our analysis of three-level systems to configurations involving at most two dissipation channels with Lindblad operators $L_1, L_2 \in \{ L_a, L_b, L_c, L_d[\alpha] \}$.
For each such configuration, determining the minimum uncertainty product $\quotient_1$ is a global optimization problem in up to $11$ real, independent parameters.
To carry out the optimizations in practice, we employed the differential evolution algorithm described in Refs.~\cite{StornJGlobalOptim1997, VirtanenNatMethods2020}.
For all configurations with two dissipation channels, we found that the minimum lies in the limit $\gamma_2 \to 0$, where the second dissipation channel is decoupled.
Like for two-level systems, the optimal three-level configuration therefore possesses only a single dissipation channel.
The minimum value that was found in our calculations,
\begin{equation}
	\quotient_{\mathrm{min}}^3 \approx 0.47242 ,
\end{equation}
is attained in a configuration where the Lindblad operator has the form $L_1 = L_d[\alpha]$.
In configurations where $L_1$ is a simple jump operator, $L_1 \in \{ L_a, L_b, L_c \}$, the minimum was approximately $0.50107$.

\begin{figure}
	\centering
	\begin{enumerate*}[\footnotesize\textbf{(\alph*)}]
	\item \adjustbox{valign=t}{
		\begin{tabular}{c|l}
			Parameter & Value \\\hline
			$n$ & $0.01096$ \\
			$\alpha$ & $0.03357$ \\
			$\bra 1 H \ket 1$ & $0$ \\
			$\bra 2 H \ket 2$ & $0$ \\
			$\bra 3 H \ket 3$ & $0$ \\
			$\bra 1 H \ket 2$ & $0.1681\, \hbar\gamma$ \\
			$\bra 2 H \ket 3$ & $0.3347\, \hbar\gamma$ \\
			$\bra 1 H \ket 3$ & $0.2164\, \ii\hbar\gamma$
		\end{tabular}}
		\quad\,
	\item
		\includegraphics[align=t,vshift=.35em]{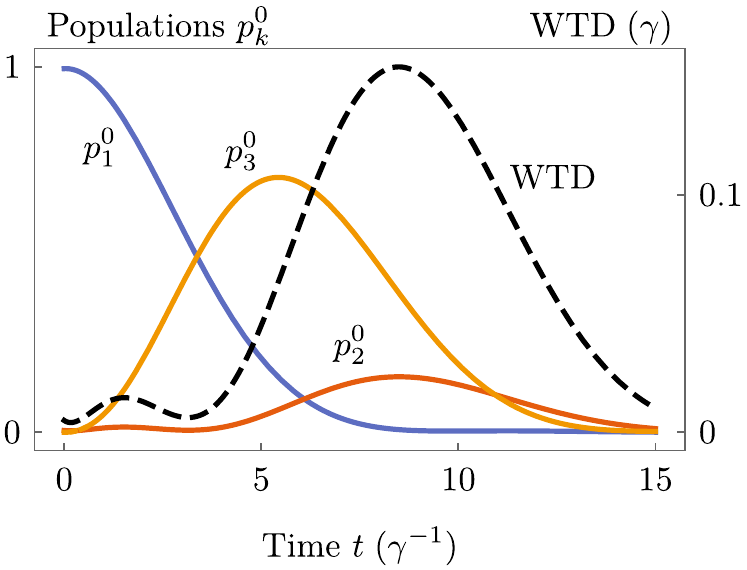}
	\end{enumerate*}
	
	\caption[.]{Three-level system with the minimum uncertainty product $\quotient_{\mathrm{min}}^3 \approx 0.47242$.
		\begin{enumerate*}[\textbf{(\alph*)}]
			\item List of the system parameters.
				Note that the rate $\gamma$ sets the time scale and can be chosen freely.
			\item Time evolution of the state after an emission event, i.e., starting from the state $\rho_0$ given in Eq.~\eqref{eq:3ls:rho0}.
				Like in Fig.~\ref{fig:2ls}, the thick solid curves show the conditional populations $p_k^0$ and are marked on the left vertical axis.
				The dashed curve shows the emission waiting time distribution and is marked on the right vertical axis.
		\end{enumerate*}}
	\label{fig:general:3ls}
\end{figure}

The time evolution of the optimal three-level system is therefore given by a master equation of the form
\begin{equation} \label{eq:3ls:lindblad}
	\partial_t \rho_t = \frac{1}{\ii\hbar} \comm{H}{\rho_t} + \gamma \mathsf D_\alpha[n] \rho_t ,
\end{equation}
where $\mathsf D_\alpha[n]$ denotes the dissipation superoperator \eqref{eq:dissipator} with Lindblad operator $L_d[\alpha]$.
The transformations $V = \sum_{j=1}^3 \e^{\ii\varphi_j} \ketbra j j$ with $\varphi_j \in \reals$ leave the form of the equation invariant and can be used to fix two phases in the system parameters; we choose $\alpha$ as well as $\bra 1 H \ket 2$ to be real and positive.
The minimum uncertainty product is then attained at a unique point in the remaining parameter space, which we specify in Fig.~\ref{fig:general:3ls}(a).
Note that all diagonal elements of $H$ are zero, generalizing the resonance condition $\delta = 0$ found for the two-level system.
In Fig.~\ref{fig:general:3ls}(b), we show -- in analogy to Fig.~\ref{fig:2ls} -- the evolution of the conditional populations $p_k^0[t]$ after an emission event at time $t=0$.
That is, we assume that the system is initially in the state
\begin{equation} \label{eq:3ls:rho0}
	\rho_0 = L_d[\alpha] \rho_\infty L_d[\alpha]^\dagger / \tr[ L_d[\alpha] \rho_\infty L_d[\alpha]^\dagger ] .
\end{equation}
Since $\alpha$ is numerically small, this state is close to the ground state $\ket 1$ and the emission and excitation probabilities are mostly proportional to the populations $p_2^0[t]$ and $p_1^0[t]$, respectively.
The plot shows that the population is first transferred from the ground state to the state $\ket 3$, which is mostly decoupled from the dissipative dynamics.
The peak of the emission waiting time distribution is therefore delayed, corresponding to smaller relative fluctuations and a lower uncertainty product.

\subsection{Four-Level Systems}

For both two- and three-level systems, we have seen that the configurations with the highest thermodynamic efficiency involve only a single dissipation channel.
For the four-level systems, we therefore focus on this type of configuration by assuming that the Lindblad equation has the form
\begin{equation} \label{eq:4ls:lindblad}
	\partial_t \rho_t = \frac{1}{\ii\hbar} \comm{H}{\rho_t} + \gamma \mathsf D_{\alpha,\beta}[n] \rho_t .
\end{equation}
Here, $\mathsf D_{\alpha,\beta}[n]$ is the dissipation superoperator for the Lindblad operator
\begin{equation}
	L[\alpha, \beta] \equiv \ketbra 1 2 + \alpha \ketbra 2 3 + \beta \ketbra 3 4 ,
\end{equation}
where $\ket j$ denotes the eigenstates of $H_0$ for $1 \leq j \leq 4$.
The symmetry of the Lindblad equation under the transformations $V = \sum_{j=1}^4 \e^{\ii\varphi_j} \ketbra j j$ for $\varphi_j \in \reals$ allows us to choose $\alpha$, $\beta$ and $\bra 1 H \ket 2$ real and positive.

\begin{figure}
	\centering
	\begin{enumerate*}[\footnotesize\textbf{(\alph*)}]
	\item \adjustbox{valign=t}{
		\begin{tabular}{c|l}
			Parameter & Value ($\hbar\gamma = 1$) \\\hline
			$n$ & $\hphantom{-}0.00560$ \\
			$\alpha$ & $\hphantom{-}0.0280$ \\
			$\beta$ & $\hphantom{-}0$ \\
			$\bra 1 H \ket 2$ & $\hphantom{-}0.101$ \\
			$\bra 2 H \ket 3$ & $\hphantom{-}0.343 - 0.025\ii$ \\
			$\bra 3 H \ket 4$ & $-0.206-0.012\ii$ \\
			$\bra 1 H \ket 3$ & $\hphantom{-}0.008 + 0.132\ii$ \\
			$\bra 2 H \ket 4$ & $-0.002 - 0.059\ii$ \\
			$\bra 1 H \ket 4$ & $\hphantom{-}0.139$
		\end{tabular}}
		\quad\,
	\item
		\includegraphics[align=t,vshift=.35em]{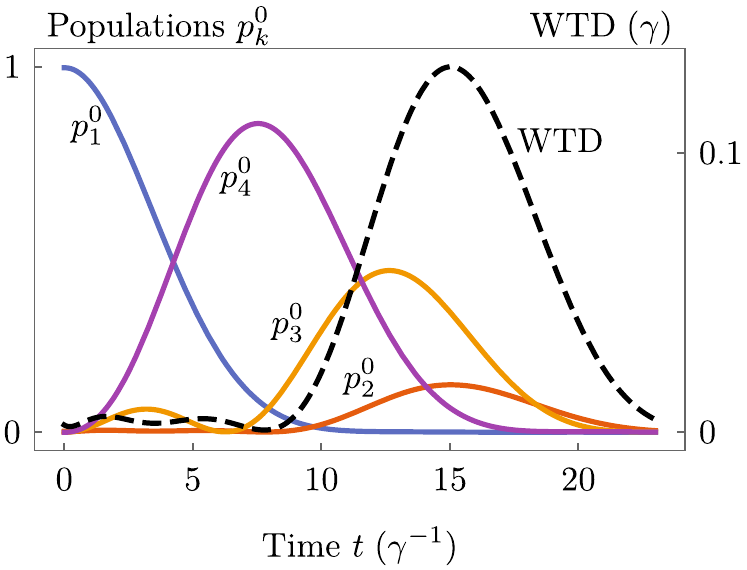}
	\end{enumerate*}
	
	\caption[.]{Four-level system with the minimum uncertainty product $\quotient_{\mathrm{min}}^4 \approx 0.2625$.
		\begin{enumerate*}[\textbf{(\alph*)}]
			\item List of the system parameters.
				Note that the matrix elements of $H$ have the dimension of energy and are listed in units of $\hbar\gamma$, which can be chosen freely.
				The diagonal matrix elements of $H$ were kept zero in the minimization.
			\item Time evolution of the state after an emission event.
				Like in Figs.~\ref{fig:2ls} and \ref{fig:general:3ls}, the thick solid curves show the conditional populations $p_k^0$ and are marked on the left vertical axis.
				The dashed curve shows the emission waiting time distribution and is marked on the right vertical axis.
		\end{enumerate*}}
	\label{fig:general:4ls}
\end{figure}

For the two- and three-level systems, we further found that the optimal configu\-rations satisfy the resonance condition
\begin{equation}
	\bra j H \ket j = 0 ,
\end{equation}
i.e., that all diagonal elements of their Hamiltonians vanish.
We assume that the same resonance condition holds for the optimal four-level system.
Determining the optimal parameter values is then a global optimization problem in $14$ real, independent parameters.
To solve it, we ran the differential evolution algorithm \cite{StornJGlobalOptim1997, VirtanenNatMethods2020} with an initial population size of $2100$ randomly selected points.
After $759$ iterations, the algorithm converged to the minimum
\begin{equation}
	\quotient_{\mathrm{min}}^4 \approx 0.2625 ,
\end{equation}
which is attained for the parameter values specified in Fig.~\ref{fig:general:4ls}(a).

The working mechanism of the optimal system, illustrated in Fig.~\ref{fig:general:4ls}(b), is similar to the three-level case.
The system state undergoes a clock-like evolution, starting close to the ground state $\ket 1$ after an emission event.
From there, the population is first transferred to the states $\ket 4$ and $\ket 3$, which couple only weakly to the environment, before it arrives in the state $\ket 2$ and can emit energy into the environment again.
The residence time of the state in the weakly coupled subspace can be made longer than for the three-level system due to the additional degree of freedom available here.
The peak of the emission waiting time distribution is therefore further delayed and the thermodynamic precision of the output heat current is larger than what is possible in three-level systems.

\section{Conclusion and Perspectives} \label{sec:conclusio}

The fact that current fluctuations on the mesoscopic and atomic scales are constrained by dissipation even far from equilibrium is surprising \emph{a priori}  \cite{HorowitzNaturePhys2020}.
It has substantial implications for the design of thermal machines, since avoiding both fluctuations and dissipation is becoming increasingly important as these machines are becoming smaller, driven by the development of quantum technologies \cite{DowlingPhilosTransRoyalSocA2003}.
This paper takes a step towards overcoming this trade-off by showing how quantum effects can alleviate such constraints.
Our analysis shows that for open quantum systems that are coupled to Markovian environments and driven into an effective non-equilibrium steady state, the thermodynamic uncertainty product is not bounded from below by $2$ like in classical systems but instead by smaller lower bounds $\quotient_{\mathrm{min}}^N$ that depend on the dimensionality $N$ of the quantum system.
The approximate values of these lower bounds as determined by our numerical investigations are
\begin{equation} \label{eq:result}
	\quotient_{\mathrm{min}}^2 \approx 1.25, \quad
	\quotient_{\mathrm{min}}^3 \approx 0.47 \quad \textup{and} \quad
	\quotient_{\mathrm{min}}^4 \approx 0.26 .
\end{equation}
Since the degrees of freedom grow rapidly with the dimension, larger system sizes are not accessible with our numerical methods.
Interestingly, the numeric value of $\quotient_{\mathrm{min}}^2$ coincides with the value of the uncertainty product in the three-qubit quantum heat engine described in Ref.~\cite{Rignon-BretPhysRevE2021}.
Deriving a general theory that provides a formula for the lower bounds $\quotient_{\mathrm{min}}^N$ is an important subject for future work and might also explain this observation.

The setups that minimize the uncertainty products in our analysis consist of a driven quantum system that is coupled to a single thermal reservoir.
Making use of coherent oscillations in the quantum system, they create clock-like cycles which convert the supplied work into heat with high thermodynamic precision.
With a larger number of accessible states, this clock-like evolution can be extended to increase the precision further.
We thus conjecture that the lower bound $\quotient_{\mathrm{min}}^N$ approaches zero as the dimension $N$ becomes large.
Our analysis of a driven cavity shows, however, that a complicated structure of the system is necessary also in high dimensions to reduce the uncertainty product below its classical bound.

The weaker constraints on fluctuations and dissipation in open quantum systems represent a genuine quantum advantage that can be exploited by quantum heat engines, as demonstrated by our three-level maser example and further examples discussed in Ref.~\cite{Rignon-BretPhysRevE2021}.
We hope that our results can be a starting point for further theoretical and experimental investigations of the role of quantum effects on the thermodynamic precision and performance of thermal machines at the nano-scale.

\ack

The research was supported by Academy of Finland through the Finnish Centre of Excellence in Quantum Technology (project nos.~312057 and 312299) and project no.~308515.
P.~M.\ acknowledges support from the Foundational Questions Institute Fund via Grant No.\ FQXi-IAF19-06.
K.~B.\ has received funding for the research presented in this paper from the Academy of Finland (Contract No.\ 296073), the University of Nottingham through a Nottingham Research Fellowship and from UK Research and Innovation through a Future Leaders Fellowship (Grant Reference: MR/S034714/1).

\appendix

\section{Spin Populations and Waiting Times} \label{app:2ls_set}

In this appendix, we summarize how the conditional populations $p^0_\pm$ and the waiting time distributions $\mathrm{WTD}$ in Fig.~\ref{fig:2ls} are calculated.
We begin with the driven \mbox{spin-$\frac 12$} system described by the Lindblad equation \eqref{eq:2ls_lindblad2}.
In order to find the conditional populations, we assume that no spontaneous emission events happen at times $t > 0$.
The time evolution of the system state is then generated by the non-trace-preserving Liouvillian
\begin{eqnarray}
	\mathsf L^0 \rho \equiv \frac{1}{\ii\hbar} \comm{\bar H + X}{\rho_t}
		&- \gamma (n + 1)\, \acomm{\sigma_+ \sigma_-}{\rho} / 2 \nonumber \\
		&+ \gamma n\, \bigl( \sigma_+ \rho \sigma_- - \acomm{\sigma_- \sigma_+}{\rho} / 2 \bigr)
\end{eqnarray}
and the conditional populations are
\begin{equation}
	p^0_\pm[t] = \bra{\pm} \Bigl( \exp\bigl[ \mathsf L^0 t \bigr]\, \ketbra - - \Bigr) \ket{\pm} ,
\end{equation}
where the time evolution superoperator $\exp[ \mathsf L^0 t ]$ acts on the ground state $\ketbra - -$.
The sum of $p^0_+[t]$ and $p^0_-[t]$ is the probability for no emission to happen until the time $t$.
The waiting time distribution is thus given by
\begin{equation} \label{eq:app:pop_wtd}
	\mathrm{WTD}[t] = -\partial_t \bigl( p^0_+[t] + p^0_-[t] \bigr) .
\end{equation}

Next, we next perform a similar analysis for the single-electron transistor.
To obtain the time evolution assuming no emission into the right lead, we modify the master equation \eqref{eq:set} into
\begin{equation}
	\partial_t \pmatrix{ {p^0_+} \cr {p^0_-} } = \gamma\, \pmatrix{ -(1 - f_L) - (1 - f_R) & +f_L + f_R \cr (1 - f_L) & -f_L - f_R } \pmatrix{ {p^0_+} \cr {p^0_-} } .
\end{equation}
Using the initial conditions $p^0_+[0] = 0$ and $p^0_-[0] = 1$, we immediately obtain the conditional populations.
The waiting time distribution can then be determined using Eq.~\eqref{eq:app:pop_wtd}.

\section{Fluctuations of the Cavity Heat Current} \label{app:ho}

In this appendix, we derive the long-time cumulants of the heat current of the driven cavity discussed in Sec.~\ref{subsec:cavity}.
We start our calculation by adding a counting field $s$ for the heat current to the Lindblad equation \eqref{eq:ho:lindblad} to obtain
\begin{eqnarray}
	\partial_t \bar\rho_t = \mathsf L[s] \bar\rho_t &\equiv -\ii\, \comm{\delta\, a^\dagger a + \Omega\, (a + a^\dagger)}{\bar\rho_t} \nonumber \\
		&\qquad + \gamma (n + 1)\, \bigl( \e^{s}\, a \bar\rho_t a^\dagger - \acomm{a^\dagger a}{\bar\rho_t} / 2 \bigr) \nonumber \\
		&\qquad + \gamma n\, \bigl( \e^{-s}\, a^\dagger \bar\rho_t a - \acomm{a a^\dagger}{\bar\rho_t} / 2 \bigr) .
\end{eqnarray}
As shown in Ref.~\cite{KubalaNewJPhys2020}, the Liouvillian $\mathsf L[s]$ is similar to the transformed Liouvillian
\begin{eqnarray}
	\mathsf L'[s] \bar\rho_t \equiv -\ii\delta\, \comm{a^\dagger a}{\bar\rho_t} &- \ii\Omega\, \bigl( \e^{-s / 2} a + \e^{s / 2} a^\dagger \bigr) \rho + \ii\Omega\, \rho \bigl( \e^{s / 2} a + \e^{-s / 2} a^\dagger \bigr) \nonumber \\
		&+ \gamma (n + 1)\, \bigl( a \bar\rho_t a^\dagger - \acomm{a^\dagger a}{\bar\rho_t} / 2 \bigr) \nonumber \\
		&+ \gamma n\, \bigl( a^\dagger \bar\rho_t a - \acomm{a a^\dagger}{\bar\rho_t} / 2 \bigr) ,
\end{eqnarray}
where the counting field now appears in the driving terms.
In other words, the superoperators $\mathsf L[s]$ and $\mathsf L'[s]$ have the same spectrum.
Since the long-time cumulants are fully determined by the spectrum of the Liouvillian, we are allowed to use $\mathsf L'[s]$ in our calculations instead of $\mathsf L[s]$.

Our goal is to determine the rescaled cumulant-generating function,
\begin{equation} \label{eq:ho:app:cgf}
	C_t = \lim_{t \to \infty} \frac 1 t \log \tr\Bigl[ \e^{\mathsf L'[s] t}\, \bar\rho_\infty \Bigr] .
\end{equation}
We recall from Eqs.~\eqref{eq:ho:stat_state} and \eqref{eq:ho:b} that the stationary state of the system is $\bar\rho_\infty = Z^{-1} \exp[ -(\hbar\omega / T)\, b^\dagger b ]$ and that $b^{(\dagger)} \equiv a^{(\dagger)} + \lambda^{(\ast)}$ are the displaced ladder operators.
We further introduce the notation
\begin{equation}
	\rgt{X} Y \equiv XY \quad \textup{and} \quad \lft{X} Y \equiv YX
\end{equation}
for any operators $X$ and $Y$.
In order to evaluate Eq.~\eqref{eq:ho:app:cgf}, we use the identity
\begin{equation} \label{eq:ho:app:identity}
	\exp\bigl[ \mathsf L'[s] t \bigr] = \mathsf{ad}[D[\alpha_t]]\, \exp\bigl[ z_t^\ast \lft{a} + z_t \rgt{a^\dagger} + x_t \bigr]\, \exp\bigl[ \mathsf L t \bigr] ,
\end{equation}
which will be proven at the end of this appendix.
Here, $\mathsf{ad}[D[\alpha_t]] \equiv \lft{D[\alpha_t]^\dagger} \rgt{D[\alpha_t]\vphantom{D^\dagger}}$ denotes conjugation with the unitary displacement operator $D[\alpha_t] \equiv \exp[ \alpha_t a^\dagger - \alpha_t^\ast a ]$ and $\mathsf L \equiv \mathsf L[0] = \mathsf L'[0]$ is the Liouvillian without the counting field.
The functions $\alpha_t$, $x_t$ and $z_t$ are given by the following expressions,
\begin{eqnarray}
	z_t &\equiv 2\lambda^\ast \sinh[s/2]\, \bigl( \e^{\gamma t / 2 - \ii\delta t} - 1 \bigr) , \nonumber \\
	\alpha_t &\equiv -(n+1)\, z_t + \bigl( 2n + 1 + \tanh[s/4] \bigr)\, z^\ast_{-t} / 2 \quad \textup{and} \nonumber \\
	x_t &\equiv \bigl( \cosh[s/2] + (2n + 1) \sinh[s/2] \bigr)\, \bigl( 2\gamma \abs{\lambda}^2 \sinh[s/2] t - \lambda^\ast z_{-t} - \lambda z^\ast_{-t} \bigr) \nonumber \\
		&\qquad + \lambda^\ast z_t + \lambda z^\ast_t + \lambda^\ast z_{-t} + \lambda z^\ast_{-t} - (n+1) \abs{z_t}^2 . \label{eq:ho:app:foo}
\end{eqnarray}
Equipped with the identity \eqref{eq:ho:app:identity}, we find
\begin{eqnarray}
	\log \tr\Bigl[ \e^{\mathsf L'[s] t}\, \bar\rho_\infty \Bigr]
		&= x_t + \log \tr\Bigl[ \e^{z_t a^\dagger}\, \bar\rho_\infty\, \e^{z_t^\ast a} \Bigr] \nonumber \\
		&= x_t - \lambda^\ast z_t - \lambda z^\ast_t + \log \int_\complex \frac{\dd^2 \beta}{\pi} \bra\beta\, e^{z_t b^\dagger}\, \bar\rho_\infty e^{z^\ast_t b} \ket\beta \nonumber \\
		&= x_t - \lambda^\ast z_t - \lambda z^\ast_t + (n+1) \abs{z_t}^2 .
\end{eqnarray}
Here, we used the overcompleteness of the coherent states $\ket\beta$ with $b\ket\beta = \beta\ket\beta$ and the Husimi representation of the equilibrium state, $\bra\beta \bar\rho_\infty \ket\beta = \exp[-\abs{\beta}^2 / (n+1)] / (n+1)$.
Plugging in Eq.~\eqref{eq:ho:app:foo} and taking the long-time average, we obtain the rescaled cumulant-generating function
\begin{equation}
	C_t = \gamma \abs{\lambda}^2 n\, \bigl(\e^s - 1\bigr) + \gamma \abs{\lambda}^2 (n+1)\, \bigl(\e^{-s} - 1\bigr) .
\end{equation}
Using $c_k \equiv (\hbar\omega)^k \partial_s^k C_t |_{s=0}$ and $c_1 = J$, Eqs.~\eqref{eq:ho:cumulants} and \eqref{eq:ho:current} immediately follow.

We still need to prove the identity \eqref{eq:ho:app:identity}.
To this end, we note that the superoperators $\mathsf L$, $\rgt{a}$, $\rgt{a^\dagger}$, $\lft{a}$, $\lft{a^\dagger}$ and $1$ form the basis of a Lie algebra $\mathcal{A}$ and that $\mathsf L'[s]$ is an element of this algebra.
Since the identity can be fully expressed in terms of commutators of members of $\mathcal{A}$, it suffices to check its validity within any faithful matrix representation.
For example, one may represent the basis of $\mathcal{A}$ as $4$-dimensional matrices as follows,
\begin{eqnarray}
\setstacktabbedgap{0.33em}
\setstackgap{L}{1.25em}
	\mathsf L \mapsto \parenMatrixstack{ 0 & -\ii\Omega & \ii\Omega & 0 \cr 0 & -\Gamma {-} \ii\delta & \gamma n & -\ii\Omega \cr 0 & -\gamma(n{+}1) & \Gamma {-} \ii\delta & -\ii\Omega \cr 0 & 0 & 0 & 0 } ,\;
	1 \mapsto \parenMatrixstack{ 0 & 0 & 0 & 1 \cr 0 & 0 & 0 & 0 \cr 0 & 0 & 0 & 0 \cr 0 & 0 & 0 & 0 } ,\;
	\lft{a^\dagger} \mapsto \parenMatrixstack{ 0 & 0 & 0 & 0 \cr 0 & 0 & 0 & 0 \cr 0 & 0 & 0 & -1 \cr 0 & 0 & 0 & 0 } , \nonumber \\
\setstacktabbedgap{0.33em}
\setstackgap{L}{1.25em}
	\rgt{ a^\dagger} \mapsto \parenMatrixstack{ 0 & 0 & 0 & 0 \cr 0 & 0 & 0 & 1 \cr 0 & 0 & 0 & 0 \cr 0 & 0 & 0 & 0 } ,\;
	\lft{a} \mapsto \parenMatrixstack{ 0 & 0 & 1 & 0 \cr 0 & 0 & 0 & 0 \cr 0 & 0 & 0 & 0 \cr 0 & 0 & 0 & 0 } \; \textup{ and } \;
	\rgt{a} \mapsto \parenMatrixstack{ 0 & 1 & 0 & 0 \cr 0 & 0 & 0 & 0 \cr 0 & 0 & 0 & 0 \cr 0 & 0 & 0 & 0 } .
\end{eqnarray}
For the sake of brevity, we have defined $\Gamma \equiv \gamma (2n + 1) / 2$.
Plugging these matrices into Eq.~\eqref{eq:ho:app:identity}, its validity can easily be verified.

\hbadness=10000
\newcommand{\newblock}{}


\begin{thebibliography}{69}%
\makeatletter
\providecommand \@ifxundefined [1]{%
 \@ifx{#1\undefined}
}%
\providecommand \@ifnum [1]{%
 \ifnum #1\expandafter \@firstoftwo
 \else \expandafter \@secondoftwo
 \fi
}%
\providecommand \@ifx [1]{%
 \ifx #1\expandafter \@firstoftwo
 \else \expandafter \@secondoftwo
 \fi
}%
\providecommand \natexlab [1]{#1}%
\providecommand \enquote  [1]{``#1''}%
\providecommand \bibnamefont  [1]{#1}%
\providecommand \bibfnamefont [1]{#1}%
\providecommand \citenamefont [1]{#1}%
\providecommand \href@noop [0]{\@secondoftwo}%
\providecommand \href [0]{\begingroup \@sanitize@url \@href}%
\providecommand \@href[1]{\@@startlink{#1}\@@href}%
\providecommand \@@href[1]{\endgroup#1\@@endlink}%
\providecommand \@sanitize@url [0]{\catcode `\\12\catcode `\$12\catcode
  `\&12\catcode `\#12\catcode `\^12\catcode `\_12\catcode `\%12\relax}%
\providecommand \@@startlink[1]{}%
\providecommand \@@endlink[0]{}%
\providecommand \url  [0]{\begingroup\@sanitize@url \@url }%
\providecommand \@url [1]{\endgroup\@href {#1}{\urlprefix }}%
\providecommand \urlprefix  [0]{URL }%
\providecommand \Eprint [0]{\href }%
\providecommand \doibase [0]{https://doi.org/}%
\providecommand \selectlanguage [0]{\@gobble}%
\providecommand \bibinfo  [0]{\@secondoftwo}%
\providecommand \bibfield  [0]{\@secondoftwo}%
\providecommand \translation [1]{[#1]}%
\providecommand \BibitemOpen [0]{}%
\providecommand \bibitemStop [0]{}%
\providecommand \bibitemNoStop [0]{.\EOS\space}%
\providecommand \EOS [0]{\spacefactor3000\relax}%
\providecommand \BibitemShut  [1]{\csname bibitem#1\endcsname}%
\let\auto@bib@innerbib\@empty
\bibitem [{\citenamefont {Manipatruni}\ \emph {et~al.}(2018)\citenamefont
  {Manipatruni}, \citenamefont {Nikonov},\ and\ \citenamefont
  {Young}}]{ManipatruniNaturePhys2018}%
  \BibitemOpen
  \bibfield  {author} {\bibinfo {author} {\bibfnamefont {S.}~\bibnamefont
  {Manipatruni}}, \bibinfo {author} {\bibfnamefont {D.~E.}\ \bibnamefont
  {Nikonov}},\ and\ \bibinfo {author} {\bibfnamefont {I.~A.}\ \bibnamefont
  {Young}},\ }\bibfield  {title} {\emph {\bibinfo {title} {Beyond {{CMOS}}
  computing with spin and polarization}},\ }\href
  {https://doi.org/10.1038/s41567-018-0101-4} {\bibfield  {journal} {\bibinfo
  {journal} {Nature Phys}\ }\textbf {\bibinfo {volume} {14}},\ \bibinfo {pages}
  {338} (\bibinfo {year} {2018})}\BibitemShut {NoStop}%
\bibitem [{\citenamefont {Kolomeisky}\ and\ \citenamefont
  {Fisher}(2007)}]{KolomeiskyAnnuRevPhysChem2007}%
  \BibitemOpen
  \bibfield  {author} {\bibinfo {author} {\bibfnamefont {A.~B.}\ \bibnamefont
  {Kolomeisky}}\ and\ \bibinfo {author} {\bibfnamefont {M.~E.}\ \bibnamefont
  {Fisher}},\ }\bibfield  {title} {\emph {\bibinfo {title} {Molecular
  {{Motors}}: {{A Theorist}}'s {{Perspective}}}},\ }\href
  {https://doi.org/10.1146/annurev.physchem.58.032806.104532} {\bibfield
  {journal} {\bibinfo  {journal} {Annu. Rev. Phys. Chem.}\ }\textbf {\bibinfo
  {volume} {58}},\ \bibinfo {pages} {675} (\bibinfo {year} {2007})}\BibitemShut
  {NoStop}%
\bibitem [{\citenamefont {Dowling}\ and\ \citenamefont
  {Milburn}(2003)}]{DowlingPhilosTransRoyalSocA2003}%
  \BibitemOpen
  \bibfield  {author} {\bibinfo {author} {\bibfnamefont {J.~P.}\ \bibnamefont
  {Dowling}}\ and\ \bibinfo {author} {\bibfnamefont {G.~J.}\ \bibnamefont
  {Milburn}},\ }\bibfield  {title} {\emph {\bibinfo {title} {Quantum
  technology: The second quantum revolution}},\ }\href
  {https://royalsocietypublishing.org/doi/abs/10.1098/rsta.2003.1227}
  {\bibfield  {journal} {\bibinfo  {journal} {Philos. Trans. Royal Soc. A}\
  }\textbf {\bibinfo {volume} {361}},\ \bibinfo {pages} {1655} (\bibinfo {year}
  {2003})}\BibitemShut {NoStop}%
\bibitem [{\citenamefont {Ladd}\ \emph {et~al.}(2010)\citenamefont {Ladd},
  \citenamefont {Jelezko}, \citenamefont {Laflamme}, \citenamefont {Nakamura},
  \citenamefont {Monroe},\ and\ \citenamefont {O'Brien}}]{LaddNature2010}%
  \BibitemOpen
  \bibfield  {author} {\bibinfo {author} {\bibfnamefont {T.~D.}\ \bibnamefont
  {Ladd}}, \bibinfo {author} {\bibfnamefont {F.}~\bibnamefont {Jelezko}},
  \bibinfo {author} {\bibfnamefont {R.}~\bibnamefont {Laflamme}}, \bibinfo
  {author} {\bibfnamefont {Y.}~\bibnamefont {Nakamura}}, \bibinfo {author}
  {\bibfnamefont {C.}~\bibnamefont {Monroe}},\ and\ \bibinfo {author}
  {\bibfnamefont {J.~L.}\ \bibnamefont {O'Brien}},\ }\bibfield  {title} {\emph
  {\bibinfo {title} {Quantum computers}},\ }\href
  {https://doi.org/10.1038/nature08812} {\bibfield  {journal} {\bibinfo
  {journal} {Nature}\ }\textbf {\bibinfo {volume} {464}},\ \bibinfo {pages}
  {45} (\bibinfo {year} {2010})}\BibitemShut {NoStop}%
\bibitem [{\citenamefont {Pekola}(2015)}]{PekolaNaturePhys2015}%
  \BibitemOpen
  \bibfield  {author} {\bibinfo {author} {\bibfnamefont {J.~P.}\ \bibnamefont
  {Pekola}},\ }\bibfield  {title} {\emph {\bibinfo {title} {Towards quantum
  thermodynamics in electronic circuits}},\ }\href
  {https://doi.org/10.1038/nphys3169} {\bibfield  {journal} {\bibinfo
  {journal} {Nature Phys}\ }\textbf {\bibinfo {volume} {11}},\ \bibinfo {pages}
  {118} (\bibinfo {year} {2015})}\BibitemShut {NoStop}%
\bibitem [{\citenamefont {Barato}\ and\ \citenamefont
  {Seifert}(2016)}]{BaratoPhysRevX2016}%
  \BibitemOpen
  \bibfield  {author} {\bibinfo {author} {\bibfnamefont {A.~C.}\ \bibnamefont
  {Barato}}\ and\ \bibinfo {author} {\bibfnamefont {U.}~\bibnamefont
  {Seifert}},\ }\bibfield  {title} {\emph {\bibinfo {title} {Cost and
  {{Precision}} of {{Brownian Clocks}}}},\ }\href
  {https://doi.org/10.1103/PhysRevX.6.041053} {\bibfield  {journal} {\bibinfo
  {journal} {Phys. Rev. X}\ }\textbf {\bibinfo {volume} {6}},\ \bibinfo {pages}
  {041053} (\bibinfo {year} {2016})}\BibitemShut {NoStop}%
\bibitem [{\citenamefont {Erker}\ \emph {et~al.}(2017)\citenamefont {Erker},
  \citenamefont {Mitchison}, \citenamefont {Silva}, \citenamefont {Woods},
  \citenamefont {Brunner},\ and\ \citenamefont {Huber}}]{ErkerPhysRevX2017}%
  \BibitemOpen
  \bibfield  {author} {\bibinfo {author} {\bibfnamefont {P.}~\bibnamefont
  {Erker}}, \bibinfo {author} {\bibfnamefont {M.~T.}\ \bibnamefont
  {Mitchison}}, \bibinfo {author} {\bibfnamefont {R.}~\bibnamefont {Silva}},
  \bibinfo {author} {\bibfnamefont {M.~P.}\ \bibnamefont {Woods}}, \bibinfo
  {author} {\bibfnamefont {N.}~\bibnamefont {Brunner}},\ and\ \bibinfo {author}
  {\bibfnamefont {M.}~\bibnamefont {Huber}},\ }\bibfield  {title} {\emph
  {\bibinfo {title} {Autonomous {{Quantum Clocks}}: {{Does Thermodynamics Limit
  Our Ability}} to {{Measure Time}}?}},\ }\href
  {https://doi.org/10.1103/PhysRevX.7.031022} {\bibfield  {journal} {\bibinfo
  {journal} {Phys. Rev. X}\ }\textbf {\bibinfo {volume} {7}},\ \bibinfo {pages}
  {031022} (\bibinfo {year} {2017})}\BibitemShut {NoStop}%
\bibitem [{\citenamefont {Mitchison}(2019)}]{MitchisonContempPhys2019}%
  \BibitemOpen
  \bibfield  {author} {\bibinfo {author} {\bibfnamefont {M.~T.}\ \bibnamefont
  {Mitchison}},\ }\bibfield  {title} {\emph {\bibinfo {title} {Quantum thermal
  absorption machines: Refrigerators, engines and clocks}},\ }\href
  {https://doi.org/10.1080/00107514.2019.1631555} {\bibfield  {journal}
  {\bibinfo  {journal} {Contemp. Phys.}\ }\textbf {\bibinfo {volume} {60}},\
  \bibinfo {pages} {164} (\bibinfo {year} {2019})}\BibitemShut {NoStop}%
\bibitem [{\citenamefont {Pietzonka}\ and\ \citenamefont
  {Seifert}(2018)}]{PietzonkaPhysRevLett2018}%
  \BibitemOpen
  \bibfield  {author} {\bibinfo {author} {\bibfnamefont {P.}~\bibnamefont
  {Pietzonka}}\ and\ \bibinfo {author} {\bibfnamefont {U.}~\bibnamefont
  {Seifert}},\ }\bibfield  {title} {\emph {\bibinfo {title} {Universal
  {{Trade}}-{{Off}} between {{Power}}, {{Efficiency}}, and {{Constancy}} in
  {{Steady}}-{{State Heat Engines}}}},\ }\href
  {https://doi.org/10.1103/PhysRevLett.120.190602} {\bibfield  {journal}
  {\bibinfo  {journal} {Phys. Rev. Lett.}\ }\textbf {\bibinfo {volume} {120}},\
  \bibinfo {pages} {190602} (\bibinfo {year} {2018})}\BibitemShut {NoStop}%
\bibitem [{\citenamefont {Barato}\ and\ \citenamefont
  {Seifert}(2015)}]{BaratoPhysRevLett2015}%
  \BibitemOpen
  \bibfield  {author} {\bibinfo {author} {\bibfnamefont {A.~C.}\ \bibnamefont
  {Barato}}\ and\ \bibinfo {author} {\bibfnamefont {U.}~\bibnamefont
  {Seifert}},\ }\bibfield  {title} {\emph {\bibinfo {title} {Thermodynamic
  {{Uncertainty Relation}} for {{Biomolecular Processes}}}},\ }\href
  {https://doi.org/10.1103/PhysRevLett.114.158101} {\bibfield  {journal}
  {\bibinfo  {journal} {Phys. Rev. Lett.}\ }\textbf {\bibinfo {volume} {114}},\
  \bibinfo {pages} {158101} (\bibinfo {year} {2015})}\BibitemShut {NoStop}%
\bibitem [{\citenamefont {Horowitz}\ and\ \citenamefont
  {Gingrich}(2020)}]{HorowitzNaturePhys2020}%
  \BibitemOpen
  \bibfield  {author} {\bibinfo {author} {\bibfnamefont {J.~M.}\ \bibnamefont
  {Horowitz}}\ and\ \bibinfo {author} {\bibfnamefont {T.~R.}\ \bibnamefont
  {Gingrich}},\ }\bibfield  {title} {\emph {\bibinfo {title} {Thermodynamic
  uncertainty relations constrain non-equilibrium fluctuations}},\ }\href
  {https://doi.org/10.1038/s41567-019-0702-6} {\bibfield  {journal} {\bibinfo
  {journal} {Nature Phys}\ }\textbf {\bibinfo {volume} {16}},\ \bibinfo {pages}
  {15} (\bibinfo {year} {2020})}\BibitemShut {NoStop}%
\bibitem [{\citenamefont {Gingrich}\ \emph {et~al.}(2016)\citenamefont
  {Gingrich}, \citenamefont {Horowitz}, \citenamefont {Perunov},\ and\
  \citenamefont {England}}]{GingrichPhysRevLett2016}%
  \BibitemOpen
  \bibfield  {author} {\bibinfo {author} {\bibfnamefont {T.~R.}\ \bibnamefont
  {Gingrich}}, \bibinfo {author} {\bibfnamefont {J.~M.}\ \bibnamefont
  {Horowitz}}, \bibinfo {author} {\bibfnamefont {N.}~\bibnamefont {Perunov}},\
  and\ \bibinfo {author} {\bibfnamefont {J.~L.}\ \bibnamefont {England}},\
  }\bibfield  {title} {\emph {\bibinfo {title} {Dissipation {{Bounds All
  Steady}}-{{State Current Fluctuations}}}},\ }\href
  {https://doi.org/10.1103/PhysRevLett.116.120601} {\bibfield  {journal}
  {\bibinfo  {journal} {Phys. Rev. Lett.}\ }\textbf {\bibinfo {volume} {116}},\
  \bibinfo {pages} {120601} (\bibinfo {year} {2016})}\BibitemShut {NoStop}%
\bibitem [{\citenamefont {Gingrich}\ \emph {et~al.}(2017)\citenamefont
  {Gingrich}, \citenamefont {Rotskoff},\ and\ \citenamefont
  {Horowitz}}]{GingrichJPhysA2017}%
  \BibitemOpen
  \bibfield  {author} {\bibinfo {author} {\bibfnamefont {T.~R.}\ \bibnamefont
  {Gingrich}}, \bibinfo {author} {\bibfnamefont {G.~M.}\ \bibnamefont
  {Rotskoff}},\ and\ \bibinfo {author} {\bibfnamefont {J.~M.}\ \bibnamefont
  {Horowitz}},\ }\bibfield  {title} {\emph {\bibinfo {title} {Inferring
  dissipation from current fluctuations}},\ }\href
  {https://doi.org/10.1088/1751-8121/aa672f} {\bibfield  {journal} {\bibinfo
  {journal} {J. Phys. A}\ }\textbf {\bibinfo {volume} {50}},\ \bibinfo {pages}
  {184004} (\bibinfo {year} {2017})}\BibitemShut {NoStop}%
\bibitem [{\citenamefont {Seifert}(2018)}]{SeifertPhysicaA2018}%
  \BibitemOpen
  \bibfield  {author} {\bibinfo {author} {\bibfnamefont {U.}~\bibnamefont
  {Seifert}},\ }\bibfield  {title} {\emph {\bibinfo {title} {Stochastic
  thermodynamics: {{From}} principles to the cost of precision}},\ }\href
  {https://doi.org/10.1016/j.physa.2017.10.024} {\bibfield  {journal} {\bibinfo
   {journal} {Physica A}\ }\textbf {\bibinfo {volume} {504}},\ \bibinfo {pages}
  {176} (\bibinfo {year} {2018})}\BibitemShut {NoStop}%
\bibitem [{\citenamefont {Dechant}\ and\ \citenamefont
  {Sasa}(2019)}]{DechantArXiv180408250Cond-Matstat-Mech2019}%
  \BibitemOpen
  \bibfield  {author} {\bibinfo {author} {\bibfnamefont {A.}~\bibnamefont
  {Dechant}}\ and\ \bibinfo {author} {\bibfnamefont {S.-i.}\ \bibnamefont
  {Sasa}},\ }\bibfield  {title} {\emph {\bibinfo {title} {Fluctuation-response
  inequality out of equilibrium}},\ }\Eprint {https://arxiv.org/abs/1804.08250}
  {arXiv:1804.08250 [cond-mat.stat-mech]}  (\bibinfo {year} {2019})\BibitemShut
  {NoStop}%
\bibitem [{\citenamefont {Potts}\ and\ \citenamefont
  {Samuelsson}(2019)}]{PottsPhysRevE2019}%
  \BibitemOpen
  \bibfield  {author} {\bibinfo {author} {\bibfnamefont {P.~P.}\ \bibnamefont
  {Potts}}\ and\ \bibinfo {author} {\bibfnamefont {P.}~\bibnamefont
  {Samuelsson}},\ }\bibfield  {title} {\emph {\bibinfo {title} {Thermodynamic
  uncertainty relations including measurement and feedback}},\ }\href
  {https://doi.org/10.1103/PhysRevE.100.052137} {\bibfield  {journal} {\bibinfo
   {journal} {Phys. Rev. E}\ }\textbf {\bibinfo {volume} {100}},\ \bibinfo
  {pages} {052137} (\bibinfo {year} {2019})}\BibitemShut {NoStop}%
\bibitem [{\citenamefont {Vu}\ and\ \citenamefont
  {Hasegawa}(2020)}]{VuJPhysA2020}%
  \BibitemOpen
  \bibfield  {author} {\bibinfo {author} {\bibfnamefont {T.~V.}\ \bibnamefont
  {Vu}}\ and\ \bibinfo {author} {\bibfnamefont {Y.}~\bibnamefont {Hasegawa}},\
  }\bibfield  {title} {\emph {\bibinfo {title} {Uncertainty relation under
  information measurement and feedback control}},\ }\href
  {https://doi.org/10.1088/1751-8121/ab64a4} {\bibfield  {journal} {\bibinfo
  {journal} {J. Phys. A}\ }\textbf {\bibinfo {volume} {53}},\ \bibinfo {pages}
  {075001} (\bibinfo {year} {2020})}\BibitemShut {NoStop}%
\bibitem [{\citenamefont {Brandner}\ \emph {et~al.}(2018)\citenamefont
  {Brandner}, \citenamefont {Hanazato},\ and\ \citenamefont
  {Saito}}]{BrandnerPhysRevLett2018}%
  \BibitemOpen
  \bibfield  {author} {\bibinfo {author} {\bibfnamefont {K.}~\bibnamefont
  {Brandner}}, \bibinfo {author} {\bibfnamefont {T.}~\bibnamefont {Hanazato}},\
  and\ \bibinfo {author} {\bibfnamefont {K.}~\bibnamefont {Saito}},\ }\bibfield
   {title} {\emph {\bibinfo {title} {Thermodynamic {{Bounds}} on {{Precision}}
  in {{Ballistic Multiterminal Transport}}}},\ }\href
  {https://doi.org/10.1103/PhysRevLett.120.090601} {\bibfield  {journal}
  {\bibinfo  {journal} {Phys. Rev. Lett.}\ }\textbf {\bibinfo {volume} {120}},\
  \bibinfo {pages} {090601} (\bibinfo {year} {2018})}\BibitemShut {NoStop}%
\bibitem [{\citenamefont {Shiraishi}\ \emph {et~al.}(2016)\citenamefont
  {Shiraishi}, \citenamefont {Saito},\ and\ \citenamefont
  {Tasaki}}]{ShiraishiPhysRevLett2016}%
  \BibitemOpen
  \bibfield  {author} {\bibinfo {author} {\bibfnamefont {N.}~\bibnamefont
  {Shiraishi}}, \bibinfo {author} {\bibfnamefont {K.}~\bibnamefont {Saito}},\
  and\ \bibinfo {author} {\bibfnamefont {H.}~\bibnamefont {Tasaki}},\
  }\bibfield  {title} {\emph {\bibinfo {title} {Universal {{Trade}}-{{Off
  Relation}} between {{Power}} and {{Efficiency}} for {{Heat Engines}}}},\
  }\href {https://doi.org/10.1103/PhysRevLett.117.190601} {\bibfield  {journal}
  {\bibinfo  {journal} {Phys. Rev. Lett.}\ }\textbf {\bibinfo {volume} {117}},\
  \bibinfo {pages} {190601} (\bibinfo {year} {2016})}\BibitemShut {NoStop}%
\bibitem [{\citenamefont {Barato}\ \emph {et~al.}(2018)\citenamefont {Barato},
  \citenamefont {Chetrite}, \citenamefont {Faggionato},\ and\ \citenamefont
  {Gabrielli}}]{BaratoNewJPhys2018}%
  \BibitemOpen
  \bibfield  {author} {\bibinfo {author} {\bibfnamefont {A.~C.}\ \bibnamefont
  {Barato}}, \bibinfo {author} {\bibfnamefont {R.}~\bibnamefont {Chetrite}},
  \bibinfo {author} {\bibfnamefont {A.}~\bibnamefont {Faggionato}},\ and\
  \bibinfo {author} {\bibfnamefont {D.}~\bibnamefont {Gabrielli}},\ }\bibfield
  {title} {\emph {\bibinfo {title} {Bounds on current fluctuations in
  periodically driven systems}},\ }\href
  {https://doi.org/10.1088/1367-2630/aae512} {\bibfield  {journal} {\bibinfo
  {journal} {New J. Phys.}\ }\textbf {\bibinfo {volume} {20}},\ \bibinfo
  {pages} {103023} (\bibinfo {year} {2018})}\BibitemShut {NoStop}%
\bibitem [{\citenamefont {Hasegawa}\ and\ \citenamefont
  {Van~Vu}(2019)}]{HasegawaPhysRevLett2019}%
  \BibitemOpen
  \bibfield  {author} {\bibinfo {author} {\bibfnamefont {Y.}~\bibnamefont
  {Hasegawa}}\ and\ \bibinfo {author} {\bibfnamefont {T.}~\bibnamefont
  {Van~Vu}},\ }\bibfield  {title} {\emph {\bibinfo {title} {Fluctuation
  {{Theorem Uncertainty Relation}}}},\ }\href
  {https://doi.org/10.1103/PhysRevLett.123.110602} {\bibfield  {journal}
  {\bibinfo  {journal} {Phys. Rev. Lett.}\ }\textbf {\bibinfo {volume} {123}},\
  \bibinfo {pages} {110602} (\bibinfo {year} {2019})}\BibitemShut {NoStop}%
\bibitem [{\citenamefont {Koyuk}\ and\ \citenamefont
  {Seifert}(2019)}]{KoyukPhysRevLett2019}%
  \BibitemOpen
  \bibfield  {author} {\bibinfo {author} {\bibfnamefont {T.}~\bibnamefont
  {Koyuk}}\ and\ \bibinfo {author} {\bibfnamefont {U.}~\bibnamefont
  {Seifert}},\ }\bibfield  {title} {\emph {\bibinfo {title} {Operationally
  {{Accessible Bounds}} on {{Fluctuations}} and {{Entropy Production}} in
  {{Periodically Driven Systems}}}},\ }\href
  {https://doi.org/10.1103/PhysRevLett.122.230601} {\bibfield  {journal}
  {\bibinfo  {journal} {Phys. Rev. Lett.}\ }\textbf {\bibinfo {volume} {122}},\
  \bibinfo {pages} {230601} (\bibinfo {year} {2019})}\BibitemShut {NoStop}%
\bibitem [{\citenamefont {Koyuk}\ and\ \citenamefont
  {Seifert}(2020)}]{KoyukPhysRevLett2020}%
  \BibitemOpen
  \bibfield  {author} {\bibinfo {author} {\bibfnamefont {T.}~\bibnamefont
  {Koyuk}}\ and\ \bibinfo {author} {\bibfnamefont {U.}~\bibnamefont
  {Seifert}},\ }\bibfield  {title} {\emph {\bibinfo {title} {Thermodynamic
  {{Uncertainty Relation}} for {{Time}}-{{Dependent Driving}}}},\ }\href
  {https://doi.org/10.1103/PhysRevLett.125.260604} {\bibfield  {journal}
  {\bibinfo  {journal} {Phys. Rev. Lett.}\ }\textbf {\bibinfo {volume} {125}},\
  \bibinfo {pages} {260604} (\bibinfo {year} {2020})}\BibitemShut {NoStop}%
\bibitem [{\citenamefont {Potanina}\ \emph {et~al.}(2021)\citenamefont
  {Potanina}, \citenamefont {Flindt}, \citenamefont {Moskalets},\ and\
  \citenamefont {Brandner}}]{PotaninaPhysRevX2021}%
  \BibitemOpen
  \bibfield  {author} {\bibinfo {author} {\bibfnamefont {E.}~\bibnamefont
  {Potanina}}, \bibinfo {author} {\bibfnamefont {C.}~\bibnamefont {Flindt}},
  \bibinfo {author} {\bibfnamefont {M.}~\bibnamefont {Moskalets}},\ and\
  \bibinfo {author} {\bibfnamefont {K.}~\bibnamefont {Brandner}},\ }\bibfield
  {title} {\emph {\bibinfo {title} {Thermodynamic bounds on coherent transport
  in periodically driven conductors}},\ }\href
  {https://doi.org/10.1103/PhysRevX.11.021013} {\bibfield  {journal} {\bibinfo
  {journal} {Phys. Rev. X}\ }\textbf {\bibinfo {volume} {11}},\ \bibinfo
  {pages} {021013} (\bibinfo {year} {2021})}\BibitemShut {NoStop}%
\bibitem [{\citenamefont {Pietzonka}\ \emph {et~al.}(2017)\citenamefont
  {Pietzonka}, \citenamefont {Ritort},\ and\ \citenamefont
  {Seifert}}]{PietzonkaPhysRevE2017}%
  \BibitemOpen
  \bibfield  {author} {\bibinfo {author} {\bibfnamefont {P.}~\bibnamefont
  {Pietzonka}}, \bibinfo {author} {\bibfnamefont {F.}~\bibnamefont {Ritort}},\
  and\ \bibinfo {author} {\bibfnamefont {U.}~\bibnamefont {Seifert}},\
  }\bibfield  {title} {\emph {\bibinfo {title} {Finite-time generalization of
  the thermodynamic uncertainty relation}},\ }\href
  {https://doi.org/10.1103/PhysRevE.96.012101} {\bibfield  {journal} {\bibinfo
  {journal} {Phys. Rev. E}\ }\textbf {\bibinfo {volume} {96}},\ \bibinfo
  {pages} {012101} (\bibinfo {year} {2017})}\BibitemShut {NoStop}%
\bibitem [{\citenamefont {Horowitz}\ and\ \citenamefont
  {Gingrich}(2017)}]{HorowitzPhysRevE2017}%
  \BibitemOpen
  \bibfield  {author} {\bibinfo {author} {\bibfnamefont {J.~M.}\ \bibnamefont
  {Horowitz}}\ and\ \bibinfo {author} {\bibfnamefont {T.~R.}\ \bibnamefont
  {Gingrich}},\ }\bibfield  {title} {\emph {\bibinfo {title} {Proof of the
  finite-time thermodynamic uncertainty relation for steady-state currents}},\
  }\href {https://doi.org/10.1103/PhysRevE.96.020103} {\bibfield  {journal}
  {\bibinfo  {journal} {Phys. Rev. E}\ }\textbf {\bibinfo {volume} {96}},\
  \bibinfo {pages} {020103} (\bibinfo {year} {2017})}\BibitemShut {NoStop}%
\bibitem [{\citenamefont {Liu}\ \emph {et~al.}(2020)\citenamefont {Liu},
  \citenamefont {Gong},\ and\ \citenamefont {Ueda}}]{LiuPhysRevLett2020}%
  \BibitemOpen
  \bibfield  {author} {\bibinfo {author} {\bibfnamefont {K.}~\bibnamefont
  {Liu}}, \bibinfo {author} {\bibfnamefont {Z.}~\bibnamefont {Gong}},\ and\
  \bibinfo {author} {\bibfnamefont {M.}~\bibnamefont {Ueda}},\ }\bibfield
  {title} {\emph {\bibinfo {title} {Thermodynamic {{Uncertainty Relation}} for
  {{Arbitrary Initial States}}}},\ }\href
  {https://doi.org/10.1103/PhysRevLett.125.140602} {\bibfield  {journal}
  {\bibinfo  {journal} {Phys. Rev. Lett.}\ }\textbf {\bibinfo {volume} {125}},\
  \bibinfo {pages} {140602} (\bibinfo {year} {2020})}\BibitemShut {NoStop}%
\bibitem [{\citenamefont {Falasco}\ \emph {et~al.}(2020)\citenamefont
  {Falasco}, \citenamefont {Esposito},\ and\ \citenamefont
  {Delvenne}}]{FalascoNewJPhys2020}%
  \BibitemOpen
  \bibfield  {author} {\bibinfo {author} {\bibfnamefont {G.}~\bibnamefont
  {Falasco}}, \bibinfo {author} {\bibfnamefont {M.}~\bibnamefont {Esposito}},\
  and\ \bibinfo {author} {\bibfnamefont {J.-C.}\ \bibnamefont {Delvenne}},\
  }\bibfield  {title} {\emph {\bibinfo {title} {Unifying thermodynamic
  uncertainty relations}},\ }\href {https://doi.org/10.1088/1367-2630/ab8679}
  {\bibfield  {journal} {\bibinfo  {journal} {New J. Phys.}\ }\textbf {\bibinfo
  {volume} {22}},\ \bibinfo {pages} {053046} (\bibinfo {year}
  {2020})}\BibitemShut {NoStop}%
\bibitem [{\citenamefont
  {Seifert}(2019)}]{SeifertAnnuRevCondensMatterPhys2019}%
  \BibitemOpen
  \bibfield  {author} {\bibinfo {author} {\bibfnamefont {U.}~\bibnamefont
  {Seifert}},\ }\bibfield  {title} {\emph {\bibinfo {title} {From {{Stochastic
  Thermodynamics}} to {{Thermodynamic Inference}}}},\ }\href
  {https://doi.org/10.1146/annurev-conmatphys-031218-013554} {\bibfield
  {journal} {\bibinfo  {journal} {Annu. Rev. Condens. Matter Phys.}\ }\textbf
  {\bibinfo {volume} {10}},\ \bibinfo {pages} {171} (\bibinfo {year}
  {2019})}\BibitemShut {NoStop}%
\bibitem [{\citenamefont {Breuer}\ and\ \citenamefont
  {Petruccione}(2002)}]{Breuer2002}%
  \BibitemOpen
  \bibfield  {author} {\bibinfo {author} {\bibfnamefont {H.-P.}\ \bibnamefont
  {Breuer}}\ and\ \bibinfo {author} {\bibfnamefont {F.}~\bibnamefont
  {Petruccione}},\ }\href@noop {} {\emph {\bibinfo {title} {The {{Theory}} of
  {{Open Quantum Systems}}}}}\ (\bibinfo  {publisher} {{Oxford University
  Press}},\ \bibinfo {address} {{Oxford}},\ \bibinfo {year} {2002})\BibitemShut
  {NoStop}%
\bibitem [{\citenamefont {Binder}\ \emph {et~al.}(2018)\citenamefont {Binder},
  \citenamefont {Correa}, \citenamefont {Gogolin}, \citenamefont {Anders},\
  and\ \citenamefont {Adesso}}]{Binder2018}%
  \BibitemOpen
  \bibinfo {editor} {\bibfnamefont {F.}~\bibnamefont {Binder}}, \bibinfo
  {editor} {\bibfnamefont {L.~A.}\ \bibnamefont {Correa}}, \bibinfo {editor}
  {\bibfnamefont {C.}~\bibnamefont {Gogolin}}, \bibinfo {editor} {\bibfnamefont
  {J.}~\bibnamefont {Anders}},\ and\ \bibinfo {editor} {\bibfnamefont
  {G.}~\bibnamefont {Adesso}},\ eds.,\ \href
  {https://doi.org/10.1007/978-3-319-99046-0} {\emph {\bibinfo {title}
  {Thermodynamics in the {{Quantum Regime}}: {{Fundamental Aspects}} and {{New
  Directions}}}}},\ \bibinfo {series} {Fundamental {{Theories}} of
  {{Physics}}}, Vol.\ \bibinfo {volume} {195}\ (\bibinfo  {publisher}
  {{Springer International Publishing}},\ \bibinfo {address} {{Cham}},\
  \bibinfo {year} {2018})\BibitemShut {NoStop}%
\bibitem [{\citenamefont {Carollo}\ \emph {et~al.}(2019)\citenamefont
  {Carollo}, \citenamefont {Jack},\ and\ \citenamefont
  {Garrahan}}]{CarolloPhysRevLett2019}%
  \BibitemOpen
  \bibfield  {author} {\bibinfo {author} {\bibfnamefont {F.}~\bibnamefont
  {Carollo}}, \bibinfo {author} {\bibfnamefont {R.~L.}\ \bibnamefont {Jack}},\
  and\ \bibinfo {author} {\bibfnamefont {J.~P.}\ \bibnamefont {Garrahan}},\
  }\bibfield  {title} {\emph {\bibinfo {title} {Unraveling the {{Large
  Deviation Statistics}} of {{Markovian Open Quantum Systems}}}},\ }\href
  {https://doi.org/10.1103/PhysRevLett.122.130605} {\bibfield  {journal}
  {\bibinfo  {journal} {Phys. Rev. Lett.}\ }\textbf {\bibinfo {volume} {122}},\
  \bibinfo {pages} {130605} (\bibinfo {year} {2019})}\BibitemShut {NoStop}%
\bibitem [{\citenamefont {Timpanaro}\ \emph {et~al.}(2019)\citenamefont
  {Timpanaro}, \citenamefont {Guarnieri}, \citenamefont {Goold},\ and\
  \citenamefont {Landi}}]{TimpanaroPhysRevLett2019}%
  \BibitemOpen
  \bibfield  {author} {\bibinfo {author} {\bibfnamefont {A.~M.}\ \bibnamefont
  {Timpanaro}}, \bibinfo {author} {\bibfnamefont {G.}~\bibnamefont
  {Guarnieri}}, \bibinfo {author} {\bibfnamefont {J.}~\bibnamefont {Goold}},\
  and\ \bibinfo {author} {\bibfnamefont {G.~T.}\ \bibnamefont {Landi}},\
  }\bibfield  {title} {\emph {\bibinfo {title} {Thermodynamic {{Uncertainty
  Relations}} from {{Exchange Fluctuation Theorems}}}},\ }\href
  {https://doi.org/10.1103/PhysRevLett.123.090604} {\bibfield  {journal}
  {\bibinfo  {journal} {Phys. Rev. Lett.}\ }\textbf {\bibinfo {volume} {123}},\
  \bibinfo {pages} {090604} (\bibinfo {year} {2019})}\BibitemShut {NoStop}%
\bibitem [{\citenamefont {Menczel}\ \emph {et~al.}(2020)\citenamefont
  {Menczel}, \citenamefont {Flindt},\ and\ \citenamefont
  {Brandner}}]{MenczelPhysRevResearch2020}%
  \BibitemOpen
  \bibfield  {author} {\bibinfo {author} {\bibfnamefont {P.}~\bibnamefont
  {Menczel}}, \bibinfo {author} {\bibfnamefont {C.}~\bibnamefont {Flindt}},\
  and\ \bibinfo {author} {\bibfnamefont {K.}~\bibnamefont {Brandner}},\
  }\bibfield  {title} {\emph {\bibinfo {title} {Quantum jump approach to
  microscopic heat engines}},\ }\href
  {https://doi.org/10.1103/PhysRevResearch.2.033449} {\bibfield  {journal}
  {\bibinfo  {journal} {Phys. Rev. Research}\ }\textbf {\bibinfo {volume}
  {2}},\ \bibinfo {pages} {033449} (\bibinfo {year} {2020})}\BibitemShut
  {NoStop}%
\bibitem [{\citenamefont {Guarnieri}\ \emph {et~al.}(2019)\citenamefont
  {Guarnieri}, \citenamefont {Landi}, \citenamefont {Clark},\ and\
  \citenamefont {Goold}}]{GuarnieriPhysRevResearch2019}%
  \BibitemOpen
  \bibfield  {author} {\bibinfo {author} {\bibfnamefont {G.}~\bibnamefont
  {Guarnieri}}, \bibinfo {author} {\bibfnamefont {G.~T.}\ \bibnamefont
  {Landi}}, \bibinfo {author} {\bibfnamefont {S.~R.}\ \bibnamefont {Clark}},\
  and\ \bibinfo {author} {\bibfnamefont {J.}~\bibnamefont {Goold}},\ }\bibfield
   {title} {\emph {\bibinfo {title} {Thermodynamics of precision in quantum
  nonequilibrium steady states}},\ }\href
  {https://doi.org/10.1103/PhysRevResearch.1.033021} {\bibfield  {journal}
  {\bibinfo  {journal} {Phys. Rev. Research}\ }\textbf {\bibinfo {volume}
  {1}},\ \bibinfo {pages} {033021} (\bibinfo {year} {2019})}\BibitemShut
  {NoStop}%
\bibitem [{\citenamefont {Hasegawa}(2020)}]{HasegawaPhysRevLett2020}%
  \BibitemOpen
  \bibfield  {author} {\bibinfo {author} {\bibfnamefont {Y.}~\bibnamefont
  {Hasegawa}},\ }\bibfield  {title} {\emph {\bibinfo {title} {Quantum
  {{Thermodynamic Uncertainty Relation}} for {{Continuous Measurement}}}},\
  }\href {https://doi.org/10.1103/PhysRevLett.125.050601} {\bibfield  {journal}
  {\bibinfo  {journal} {Phys. Rev. Lett.}\ }\textbf {\bibinfo {volume} {125}},\
  \bibinfo {pages} {050601} (\bibinfo {year} {2020})}\BibitemShut {NoStop}%
\bibitem [{\citenamefont {Hasegawa}(2021)}]{HasegawaPhysRevLett2021}%
  \BibitemOpen
  \bibfield  {author} {\bibinfo {author} {\bibfnamefont {Y.}~\bibnamefont
  {Hasegawa}},\ }\bibfield  {title} {\emph {\bibinfo {title} {Thermodynamic
  {{Uncertainty Relation}} for {{General Open Quantum Systems}}}},\ }\href
  {https://doi.org/10.1103/PhysRevLett.126.010602} {\bibfield  {journal}
  {\bibinfo  {journal} {Phys. Rev. Lett.}\ }\textbf {\bibinfo {volume} {126}},\
  \bibinfo {pages} {010602} (\bibinfo {year} {2021})}\BibitemShut {NoStop}%
\bibitem [{\citenamefont {{Rignon-Bret}}\ \emph {et~al.}(2021)\citenamefont
  {{Rignon-Bret}}, \citenamefont {Guarnieri}, \citenamefont {Goold},\ and\
  \citenamefont {Mitchison}}]{Rignon-BretPhysRevE2021}%
  \BibitemOpen
  \bibfield  {author} {\bibinfo {author} {\bibfnamefont {A.}~\bibnamefont
  {{Rignon-Bret}}}, \bibinfo {author} {\bibfnamefont {G.}~\bibnamefont
  {Guarnieri}}, \bibinfo {author} {\bibfnamefont {J.}~\bibnamefont {Goold}},\
  and\ \bibinfo {author} {\bibfnamefont {M.~T.}\ \bibnamefont {Mitchison}},\
  }\bibfield  {title} {\emph {\bibinfo {title} {Thermodynamics of precision in
  quantum nanomachines}},\ }\href {https://doi.org/10.1103/PhysRevE.103.012133}
  {\bibfield  {journal} {\bibinfo  {journal} {Phys. Rev. E}\ }\textbf {\bibinfo
  {volume} {103}},\ \bibinfo {pages} {012133} (\bibinfo {year}
  {2021})}\BibitemShut {NoStop}%
\bibitem [{\citenamefont {Miller}\ \emph {et~al.}(2021)\citenamefont {Miller},
  \citenamefont {Mohammady}, \citenamefont {{Perarnau-Llobet}},\ and\
  \citenamefont {Guarnieri}}]{MillerPhysRevLett2021}%
  \BibitemOpen
  \bibfield  {author} {\bibinfo {author} {\bibfnamefont {H.~J.~D.}\
  \bibnamefont {Miller}}, \bibinfo {author} {\bibfnamefont {M.~H.}\
  \bibnamefont {Mohammady}}, \bibinfo {author} {\bibfnamefont {M.}~\bibnamefont
  {{Perarnau-Llobet}}},\ and\ \bibinfo {author} {\bibfnamefont
  {G.}~\bibnamefont {Guarnieri}},\ }\bibfield  {title} {\emph {\bibinfo {title}
  {Thermodynamic {{Uncertainty Relation}} in {{Slowly Driven Quantum Heat
  Engines}}}},\ }\href {https://doi.org/10.1103/PhysRevLett.126.210603}
  {\bibfield  {journal} {\bibinfo  {journal} {Phys. Rev. Lett.}\ }\textbf
  {\bibinfo {volume} {126}},\ \bibinfo {pages} {210603} (\bibinfo {year}
  {2021})}\BibitemShut {NoStop}%
\bibitem [{\citenamefont {Tajima}\ and\ \citenamefont
  {Funo}(2020)}]{TajimaArXiv200413412Quant-Ph2020}%
  \BibitemOpen
  \bibfield  {author} {\bibinfo {author} {\bibfnamefont {H.}~\bibnamefont
  {Tajima}}\ and\ \bibinfo {author} {\bibfnamefont {K.}~\bibnamefont {Funo}},\
  }\bibfield  {title} {\emph {\bibinfo {title} {Superconducting-like heat
  current: {{Effective}} cancellation of current-dissipation trade off by
  quantum coherence}},\ }\Eprint {https://arxiv.org/abs/2004.13412}
  {arXiv:2004.13412 [quant-ph]}  (\bibinfo {year} {2020})\BibitemShut {NoStop}%
\bibitem [{\citenamefont {Le~Bellac}(2006)}]{LeBellac2006}%
  \BibitemOpen
  \bibfield  {author} {\bibinfo {author} {\bibfnamefont {M.}~\bibnamefont
  {Le~Bellac}},\ }\href@noop {} {\emph {\bibinfo {title} {A {{Short
  Introduction}} to {{Quantum Information}} and {{Quantum Computation}}}}}\
  (\bibinfo  {publisher} {{Cambridge University Press}},\ \bibinfo {address}
  {{Cambridge}},\ \bibinfo {year} {2006})\BibitemShut {NoStop}%
\bibitem [{\citenamefont {Alicki}(1979)}]{AlickiJPhysA1979}%
  \BibitemOpen
  \bibfield  {author} {\bibinfo {author} {\bibfnamefont {R.}~\bibnamefont
  {Alicki}},\ }\bibfield  {title} {\emph {\bibinfo {title} {The quantum open
  system as a model of the heat engine}},\ }\href
  {https://doi.org/10.1088/0305-4470/12/5/007} {\bibfield  {journal} {\bibinfo
  {journal} {J. Phys. A}\ }\textbf {\bibinfo {volume} {12}},\ \bibinfo {pages}
  {L103} (\bibinfo {year} {1979})}\BibitemShut {NoStop}%
\bibitem [{\citenamefont {Albash}\ \emph {et~al.}(2012)\citenamefont {Albash},
  \citenamefont {Boixo}, \citenamefont {Lidar},\ and\ \citenamefont
  {Zanardi}}]{AlbashNewJPhys2012}%
  \BibitemOpen
  \bibfield  {author} {\bibinfo {author} {\bibfnamefont {T.}~\bibnamefont
  {Albash}}, \bibinfo {author} {\bibfnamefont {S.}~\bibnamefont {Boixo}},
  \bibinfo {author} {\bibfnamefont {D.~A.}\ \bibnamefont {Lidar}},\ and\
  \bibinfo {author} {\bibfnamefont {P.}~\bibnamefont {Zanardi}},\ }\bibfield
  {title} {\emph {\bibinfo {title} {Quantum adiabatic {{Markovian}} master
  equations}},\ }\href {https://doi.org/10.1088/1367-2630/14/12/123016}
  {\bibfield  {journal} {\bibinfo  {journal} {New J. Phys.}\ }\textbf {\bibinfo
  {volume} {14}},\ \bibinfo {pages} {123016} (\bibinfo {year}
  {2012})}\BibitemShut {NoStop}%
\bibitem [{\citenamefont {Brandner}\ and\ \citenamefont
  {Saito}(2020)}]{BrandnerPhysRevLett2020}%
  \BibitemOpen
  \bibfield  {author} {\bibinfo {author} {\bibfnamefont {K.}~\bibnamefont
  {Brandner}}\ and\ \bibinfo {author} {\bibfnamefont {K.}~\bibnamefont
  {Saito}},\ }\bibfield  {title} {\emph {\bibinfo {title} {Thermodynamic
  {{Geometry}} of {{Microscopic Heat Engines}}}},\ }\href
  {https://doi.org/10.1103/PhysRevLett.124.040602} {\bibfield  {journal}
  {\bibinfo  {journal} {Phys. Rev. Lett.}\ }\textbf {\bibinfo {volume} {124}},\
  \bibinfo {pages} {040602} (\bibinfo {year} {2020})}\BibitemShut {NoStop}%
\bibitem [{\citenamefont {Gorini}\ \emph {et~al.}(1978)\citenamefont {Gorini},
  \citenamefont {Frigerio}, \citenamefont {Verri}, \citenamefont
  {Kossakowski},\ and\ \citenamefont {Sudarshan}}]{GoriniRepMathPhys1978}%
  \BibitemOpen
  \bibfield  {author} {\bibinfo {author} {\bibfnamefont {V.}~\bibnamefont
  {Gorini}}, \bibinfo {author} {\bibfnamefont {A.}~\bibnamefont {Frigerio}},
  \bibinfo {author} {\bibfnamefont {M.}~\bibnamefont {Verri}}, \bibinfo
  {author} {\bibfnamefont {A.}~\bibnamefont {Kossakowski}},\ and\ \bibinfo
  {author} {\bibfnamefont {E.~C.~G.}\ \bibnamefont {Sudarshan}},\ }\bibfield
  {title} {\emph {\bibinfo {title} {Properties of quantum {{Markovian}} master
  equations}},\ }\href {https://doi.org/10.1016/0034-4877(78)90050-2}
  {\bibfield  {journal} {\bibinfo  {journal} {Rep. Math. Phys.}\ }\textbf
  {\bibinfo {volume} {13}},\ \bibinfo {pages} {149} (\bibinfo {year}
  {1978})}\BibitemShut {NoStop}%
\bibitem [{\citenamefont {Brandner}\ and\ \citenamefont
  {Seifert}(2016)}]{BrandnerPhysRevE2016}%
  \BibitemOpen
  \bibfield  {author} {\bibinfo {author} {\bibfnamefont {K.}~\bibnamefont
  {Brandner}}\ and\ \bibinfo {author} {\bibfnamefont {U.}~\bibnamefont
  {Seifert}},\ }\bibfield  {title} {\emph {\bibinfo {title} {Periodic
  thermodynamics of open quantum systems}},\ }\href
  {https://doi.org/10.1103/PhysRevE.93.062134} {\bibfield  {journal} {\bibinfo
  {journal} {Phys. Rev. E}\ }\textbf {\bibinfo {volume} {93}},\ \bibinfo
  {pages} {062134} (\bibinfo {year} {2016})}\BibitemShut {NoStop}%
\bibitem [{\citenamefont {Spohn}(1977)}]{SpohnLettMathPhys1977}%
  \BibitemOpen
  \bibfield  {author} {\bibinfo {author} {\bibfnamefont {H.}~\bibnamefont
  {Spohn}},\ }\bibfield  {title} {\emph {\bibinfo {title} {An algebraic
  condition for the approach to equilibrium of an open {{N}}-level system}},\
  }\href {https://doi.org/10.1007/BF00420668} {\bibfield  {journal} {\bibinfo
  {journal} {Lett. Math. Phys}\ }\textbf {\bibinfo {volume} {2}},\ \bibinfo
  {pages} {33} (\bibinfo {year} {1977})}\BibitemShut {NoStop}%
\bibitem [{\citenamefont {Menczel}\ and\ \citenamefont
  {Brandner}(2019)}]{MenczelJPhysA2019}%
  \BibitemOpen
  \bibfield  {author} {\bibinfo {author} {\bibfnamefont {P.}~\bibnamefont
  {Menczel}}\ and\ \bibinfo {author} {\bibfnamefont {K.}~\bibnamefont
  {Brandner}},\ }\bibfield  {title} {\emph {\bibinfo {title} {Limit cycles in
  periodically driven open quantum systems}},\ }\href
  {https://doi.org/10.1088/1751-8121/ab435a} {\bibfield  {journal} {\bibinfo
  {journal} {J. Phys. A}\ }\textbf {\bibinfo {volume} {52}},\ \bibinfo {pages}
  {43LT01} (\bibinfo {year} {2019})}\BibitemShut {NoStop}%
\bibitem [{\citenamefont {Levitov}\ \emph {et~al.}(1996)\citenamefont
  {Levitov}, \citenamefont {Lee},\ and\ \citenamefont
  {Lesovik}}]{LevitovJMathPhys1996}%
  \BibitemOpen
  \bibfield  {author} {\bibinfo {author} {\bibfnamefont {L.~S.}\ \bibnamefont
  {Levitov}}, \bibinfo {author} {\bibfnamefont {H.}~\bibnamefont {Lee}},\ and\
  \bibinfo {author} {\bibfnamefont {G.~B.}\ \bibnamefont {Lesovik}},\
  }\bibfield  {title} {\emph {\bibinfo {title} {Electron counting statistics
  and coherent states of electric current}},\ }\href
  {https://doi.org/10.1063/1.531672} {\bibfield  {journal} {\bibinfo  {journal}
  {J. Math. Phys.}\ }\textbf {\bibinfo {volume} {37}},\ \bibinfo {pages} {4845}
  (\bibinfo {year} {1996})}\BibitemShut {NoStop}%
\bibitem [{\citenamefont {Bagrets}\ and\ \citenamefont
  {Nazarov}(2003)}]{BagretsPhysRevB2003}%
  \BibitemOpen
  \bibfield  {author} {\bibinfo {author} {\bibfnamefont {D.~A.}\ \bibnamefont
  {Bagrets}}\ and\ \bibinfo {author} {\bibfnamefont {Y.~V.}\ \bibnamefont
  {Nazarov}},\ }\bibfield  {title} {\emph {\bibinfo {title} {Full counting
  statistics of charge transfer in {{Coulomb}} blockade systems}},\ }\href
  {https://doi.org/10.1103/PhysRevB.67.085316} {\bibfield  {journal} {\bibinfo
  {journal} {Phys. Rev. B}\ }\textbf {\bibinfo {volume} {67}},\ \bibinfo
  {pages} {085316} (\bibinfo {year} {2003})}\BibitemShut {NoStop}%
\bibitem [{\citenamefont {Flindt}\ \emph {et~al.}(2004)\citenamefont {Flindt},
  \citenamefont {Novotn{\'y}},\ and\ \citenamefont {Jauho}}]{FlindtEPL2004}%
  \BibitemOpen
  \bibfield  {author} {\bibinfo {author} {\bibfnamefont {C.}~\bibnamefont
  {Flindt}}, \bibinfo {author} {\bibfnamefont {T.}~\bibnamefont
  {Novotn{\'y}}},\ and\ \bibinfo {author} {\bibfnamefont {A.-P.}\ \bibnamefont
  {Jauho}},\ }\bibfield  {title} {\emph {\bibinfo {title} {Full counting
  statistics of nano-electromechanical systems}},\ }\href
  {https://doi.org/10.1209/epl/i2004-10351-x} {\bibfield  {journal} {\bibinfo
  {journal} {EPL}\ }\textbf {\bibinfo {volume} {69}},\ \bibinfo {pages} {475}
  (\bibinfo {year} {2004})}\BibitemShut {NoStop}%
\bibitem [{\citenamefont {Bruderer}\ \emph {et~al.}(2014)\citenamefont
  {Bruderer}, \citenamefont {{Contreras-Pulido}}, \citenamefont {Thaller},
  \citenamefont {Sironi}, \citenamefont {Obreschkow},\ and\ \citenamefont
  {Plenio}}]{BrudererNewJPhys2014}%
  \BibitemOpen
  \bibfield  {author} {\bibinfo {author} {\bibfnamefont {M.}~\bibnamefont
  {Bruderer}}, \bibinfo {author} {\bibfnamefont {L.~D.}\ \bibnamefont
  {{Contreras-Pulido}}}, \bibinfo {author} {\bibfnamefont {M.}~\bibnamefont
  {Thaller}}, \bibinfo {author} {\bibfnamefont {L.}~\bibnamefont {Sironi}},
  \bibinfo {author} {\bibfnamefont {D.}~\bibnamefont {Obreschkow}},\ and\
  \bibinfo {author} {\bibfnamefont {M.~B.}\ \bibnamefont {Plenio}},\ }\bibfield
   {title} {\emph {\bibinfo {title} {Inverse counting statistics for stochastic
  and open quantum systems: The characteristic polynomial approach}},\ }\href
  {https://doi.org/10.1088/1367-2630/16/3/033030} {\bibfield  {journal}
  {\bibinfo  {journal} {New J. Phys.}\ }\textbf {\bibinfo {volume} {16}},\
  \bibinfo {pages} {033030} (\bibinfo {year} {2014})}\BibitemShut {NoStop}%
\bibitem [{\citenamefont {Plenio}\ and\ \citenamefont
  {Knight}(1998)}]{PlenioRevModPhys1998}%
  \BibitemOpen
  \bibfield  {author} {\bibinfo {author} {\bibfnamefont {M.~B.}\ \bibnamefont
  {Plenio}}\ and\ \bibinfo {author} {\bibfnamefont {P.~L.}\ \bibnamefont
  {Knight}},\ }\bibfield  {title} {\emph {\bibinfo {title} {The quantum-jump
  approach to dissipative dynamics in quantum optics}},\ }\href
  {https://doi.org/10.1103/RevModPhys.70.101} {\bibfield  {journal} {\bibinfo
  {journal} {Rev. Mod. Phys.}\ }\textbf {\bibinfo {volume} {70}},\ \bibinfo
  {pages} {101} (\bibinfo {year} {1998})}\BibitemShut {NoStop}%
\bibitem [{\citenamefont {Carmichael}\ \emph {et~al.}(1989)\citenamefont
  {Carmichael}, \citenamefont {Singh}, \citenamefont {Vyas},\ and\
  \citenamefont {Rice}}]{CarmichaelPhysRevA1989}%
  \BibitemOpen
  \bibfield  {author} {\bibinfo {author} {\bibfnamefont {H.~J.}\ \bibnamefont
  {Carmichael}}, \bibinfo {author} {\bibfnamefont {S.}~\bibnamefont {Singh}},
  \bibinfo {author} {\bibfnamefont {R.}~\bibnamefont {Vyas}},\ and\ \bibinfo
  {author} {\bibfnamefont {P.~R.}\ \bibnamefont {Rice}},\ }\bibfield  {title}
  {\emph {\bibinfo {title} {Photoelectron waiting times and atomic state
  reduction in resonance fluorescence}},\ }\href
  {https://doi.org/10.1103/PhysRevA.39.1200} {\bibfield  {journal} {\bibinfo
  {journal} {Phys. Rev. A}\ }\textbf {\bibinfo {volume} {39}},\ \bibinfo
  {pages} {1200} (\bibinfo {year} {1989})}\BibitemShut {NoStop}%
\bibitem [{\citenamefont {Albert}\ \emph {et~al.}(2012)\citenamefont {Albert},
  \citenamefont {Haack}, \citenamefont {Flindt},\ and\ \citenamefont
  {B{\"u}ttiker}}]{AlbertPhysRevLett2012}%
  \BibitemOpen
  \bibfield  {author} {\bibinfo {author} {\bibfnamefont {M.}~\bibnamefont
  {Albert}}, \bibinfo {author} {\bibfnamefont {G.}~\bibnamefont {Haack}},
  \bibinfo {author} {\bibfnamefont {C.}~\bibnamefont {Flindt}},\ and\ \bibinfo
  {author} {\bibfnamefont {M.}~\bibnamefont {B{\"u}ttiker}},\ }\bibfield
  {title} {\emph {\bibinfo {title} {Electron {{Waiting Times}} in {{Mesoscopic
  Conductors}}}},\ }\href {https://doi.org/10.1103/PhysRevLett.108.186806}
  {\bibfield  {journal} {\bibinfo  {journal} {Phys. Rev. Lett.}\ }\textbf
  {\bibinfo {volume} {108}},\ \bibinfo {pages} {186806} (\bibinfo {year}
  {2012})}\BibitemShut {NoStop}%
\bibitem [{\citenamefont {Haack}\ \emph {et~al.}(2014)\citenamefont {Haack},
  \citenamefont {Albert},\ and\ \citenamefont {Flindt}}]{HaackPhysRevB2014}%
  \BibitemOpen
  \bibfield  {author} {\bibinfo {author} {\bibfnamefont {G.}~\bibnamefont
  {Haack}}, \bibinfo {author} {\bibfnamefont {M.}~\bibnamefont {Albert}},\ and\
  \bibinfo {author} {\bibfnamefont {C.}~\bibnamefont {Flindt}},\ }\bibfield
  {title} {\emph {\bibinfo {title} {Distributions of electron waiting times in
  quantum-coherent conductors}},\ }\href
  {https://doi.org/10.1103/PhysRevB.90.205429} {\bibfield  {journal} {\bibinfo
  {journal} {Phys. Rev. B}\ }\textbf {\bibinfo {volume} {90}},\ \bibinfo
  {pages} {205429} (\bibinfo {year} {2014})}\BibitemShut {NoStop}%
\bibitem [{\citenamefont {Schaller}(2015)}]{Schaller2015}%
  \BibitemOpen
  \bibfield  {author} {\bibinfo {author} {\bibfnamefont {G.}~\bibnamefont
  {Schaller}},\ }\href@noop {} {\emph {\bibinfo {title} {Non-{{Equilibrium
  Master Equations}}}}}\ (\bibinfo {year} {2015})\BibitemShut {NoStop}%
\bibitem [{\citenamefont {Carmichael}(2008)}]{Carmichael2008}%
  \BibitemOpen
  \bibfield  {author} {\bibinfo {author} {\bibfnamefont {H.~J.}\ \bibnamefont
  {Carmichael}},\ }\href {https://doi.org/10.1007/978-3-540-71320-3} {\emph
  {\bibinfo {title} {Statistical {{Methods}} in {{Quantum Optics}} 2}}},\
  Theoretical and {{Mathematical Physics}}\ (\bibinfo  {publisher}
  {{Springer-Verlag}},\ \bibinfo {address} {{Berlin Heidelberg}},\ \bibinfo
  {year} {2008})\BibitemShut {NoStop}%
\bibitem [{\citenamefont {Hofer}(2016)}]{Hofer2016}%
  \BibitemOpen
  \bibfield  {author} {\bibinfo {author} {\bibfnamefont {P.}~\bibnamefont
  {Hofer}},\ }\emph {\bibinfo {title} {Dynamic Mesoscopic Conductors: Single
  Electron Sources, Full Counting Statistics, and Thermal Machines}},\
  \href@noop {} {Ph.D. thesis} (\bibinfo {year} {2016})\BibitemShut {NoStop}%
\bibitem [{\citenamefont {Scovil}\ and\ \citenamefont
  {{Schulz-DuBois}}(1959)}]{ScovilPhysRevLett1959}%
  \BibitemOpen
  \bibfield  {author} {\bibinfo {author} {\bibfnamefont {H.~E.~D.}\
  \bibnamefont {Scovil}}\ and\ \bibinfo {author} {\bibfnamefont {E.~O.}\
  \bibnamefont {{Schulz-DuBois}}},\ }\bibfield  {title} {\emph {\bibinfo
  {title} {Three-{{Level Masers}} as {{Heat Engines}}}},\ }\href
  {https://doi.org/10.1103/PhysRevLett.2.262} {\bibfield  {journal} {\bibinfo
  {journal} {Phys. Rev. Lett.}\ }\textbf {\bibinfo {volume} {2}},\ \bibinfo
  {pages} {262} (\bibinfo {year} {1959})}\BibitemShut {NoStop}%
\bibitem [{\citenamefont {Zou}\ \emph {et~al.}(2017)\citenamefont {Zou},
  \citenamefont {Jiang}, \citenamefont {Mei}, \citenamefont {Guo},\ and\
  \citenamefont {Du}}]{ZouPhysRevLett2017}%
  \BibitemOpen
  \bibfield  {author} {\bibinfo {author} {\bibfnamefont {Y.}~\bibnamefont
  {Zou}}, \bibinfo {author} {\bibfnamefont {Y.}~\bibnamefont {Jiang}}, \bibinfo
  {author} {\bibfnamefont {Y.}~\bibnamefont {Mei}}, \bibinfo {author}
  {\bibfnamefont {X.}~\bibnamefont {Guo}},\ and\ \bibinfo {author}
  {\bibfnamefont {S.}~\bibnamefont {Du}},\ }\bibfield  {title} {\emph {\bibinfo
  {title} {Quantum {{Heat Engine Using Electromagnetically Induced
  Transparency}}}},\ }\href {https://doi.org/10.1103/PhysRevLett.119.050602}
  {\bibfield  {journal} {\bibinfo  {journal} {Phys. Rev. Lett.}\ }\textbf
  {\bibinfo {volume} {119}},\ \bibinfo {pages} {050602} (\bibinfo {year}
  {2017})}\BibitemShut {NoStop}%
\bibitem [{\citenamefont {Klatzow}\ \emph {et~al.}(2019)\citenamefont
  {Klatzow}, \citenamefont {Becker}, \citenamefont {Ledingham}, \citenamefont
  {Weinzetl}, \citenamefont {Kaczmarek}, \citenamefont {Saunders},
  \citenamefont {Nunn}, \citenamefont {Walmsley}, \citenamefont {Uzdin},\ and\
  \citenamefont {Poem}}]{KlatzowPhysRevLett2019}%
  \BibitemOpen
  \bibfield  {author} {\bibinfo {author} {\bibfnamefont {J.}~\bibnamefont
  {Klatzow}}, \bibinfo {author} {\bibfnamefont {J.~N.}\ \bibnamefont {Becker}},
  \bibinfo {author} {\bibfnamefont {P.~M.}\ \bibnamefont {Ledingham}}, \bibinfo
  {author} {\bibfnamefont {C.}~\bibnamefont {Weinzetl}}, \bibinfo {author}
  {\bibfnamefont {K.~T.}\ \bibnamefont {Kaczmarek}}, \bibinfo {author}
  {\bibfnamefont {D.~J.}\ \bibnamefont {Saunders}}, \bibinfo {author}
  {\bibfnamefont {J.}~\bibnamefont {Nunn}}, \bibinfo {author} {\bibfnamefont
  {I.~A.}\ \bibnamefont {Walmsley}}, \bibinfo {author} {\bibfnamefont
  {R.}~\bibnamefont {Uzdin}},\ and\ \bibinfo {author} {\bibfnamefont
  {E.}~\bibnamefont {Poem}},\ }\bibfield  {title} {\emph {\bibinfo {title}
  {Experimental {{Demonstration}} of {{Quantum Effects}} in the {{Operation}}
  of {{Microscopic Heat Engines}}}},\ }\href
  {https://doi.org/10.1103/PhysRevLett.122.110601} {\bibfield  {journal}
  {\bibinfo  {journal} {Phys. Rev. Lett.}\ }\textbf {\bibinfo {volume} {122}},\
  \bibinfo {pages} {110601} (\bibinfo {year} {2019})}\BibitemShut {NoStop}%
\bibitem [{\citenamefont {Rivas}\ \emph {et~al.}(2010)\citenamefont {Rivas},
  \citenamefont {Plato}, \citenamefont {Huelga},\ and\ \citenamefont
  {Plenio}}]{RivasNewJPhys2010}%
  \BibitemOpen
  \bibfield  {author} {\bibinfo {author} {\bibfnamefont {{\'A}.}~\bibnamefont
  {Rivas}}, \bibinfo {author} {\bibfnamefont {A.~D.~K.}\ \bibnamefont {Plato}},
  \bibinfo {author} {\bibfnamefont {S.~F.}\ \bibnamefont {Huelga}},\ and\
  \bibinfo {author} {\bibfnamefont {M.~B.}\ \bibnamefont {Plenio}},\ }\bibfield
   {title} {\emph {\bibinfo {title} {Markovian master equations: A critical
  study}},\ }\href {https://doi.org/10.1088/1367-2630/12/11/113032} {\bibfield
  {journal} {\bibinfo  {journal} {New J. Phys.}\ }\textbf {\bibinfo {volume}
  {12}},\ \bibinfo {pages} {113032} (\bibinfo {year} {2010})}\BibitemShut
  {NoStop}%
\bibitem [{\citenamefont {Prosen}(2011)}]{ProsenPhysRevLett2011}%
  \BibitemOpen
  \bibfield  {author} {\bibinfo {author} {\bibfnamefont {T.}~\bibnamefont
  {Prosen}},\ }\bibfield  {title} {\emph {\bibinfo {title} {Exact
  {{Nonequilibrium Steady State}} of a {{Strongly Driven Open}} {$XXZ$}
  {{Chain}}}},\ }\href {https://doi.org/10.1103/PhysRevLett.107.137201}
  {\bibfield  {journal} {\bibinfo  {journal} {Phys. Rev. Lett.}\ }\textbf
  {\bibinfo {volume} {107}},\ \bibinfo {pages} {137201} (\bibinfo {year}
  {2011})}\BibitemShut {NoStop}%
\bibitem [{\citenamefont {Karevski}\ \emph {et~al.}(2013)\citenamefont
  {Karevski}, \citenamefont {Popkov},\ and\ \citenamefont
  {Sch{\"u}tz}}]{KarevskiPhysRevLett2013}%
  \BibitemOpen
  \bibfield  {author} {\bibinfo {author} {\bibfnamefont {D.}~\bibnamefont
  {Karevski}}, \bibinfo {author} {\bibfnamefont {V.}~\bibnamefont {Popkov}},\
  and\ \bibinfo {author} {\bibfnamefont {G.~M.}\ \bibnamefont {Sch{\"u}tz}},\
  }\bibfield  {title} {\emph {\bibinfo {title} {Exact {{Matrix Product
  Solution}} for the {{Boundary}}-{{Driven Lindblad}} {$XXZ$} {{Chain}}}},\
  }\href {https://doi.org/10.1103/PhysRevLett.110.047201} {\bibfield  {journal}
  {\bibinfo  {journal} {Phys. Rev. Lett.}\ }\textbf {\bibinfo {volume} {110}},\
  \bibinfo {pages} {047201} (\bibinfo {year} {2013})}\BibitemShut {NoStop}%
\bibitem [{\citenamefont {Chiara}\ \emph {et~al.}(2018)\citenamefont {Chiara},
  \citenamefont {Landi}, \citenamefont {Hewgill}, \citenamefont {Reid},
  \citenamefont {Ferraro}, \citenamefont {Roncaglia},\ and\ \citenamefont
  {Antezza}}]{ChiaraNewJPhys2018}%
  \BibitemOpen
  \bibfield  {author} {\bibinfo {author} {\bibfnamefont {G.~D.}\ \bibnamefont
  {Chiara}}, \bibinfo {author} {\bibfnamefont {G.}~\bibnamefont {Landi}},
  \bibinfo {author} {\bibfnamefont {A.}~\bibnamefont {Hewgill}}, \bibinfo
  {author} {\bibfnamefont {B.}~\bibnamefont {Reid}}, \bibinfo {author}
  {\bibfnamefont {A.}~\bibnamefont {Ferraro}}, \bibinfo {author} {\bibfnamefont
  {A.~J.}\ \bibnamefont {Roncaglia}},\ and\ \bibinfo {author} {\bibfnamefont
  {M.}~\bibnamefont {Antezza}},\ }\bibfield  {title} {\emph {\bibinfo {title}
  {Reconciliation of quantum local master equations with thermodynamics}},\
  }\href {https://doi.org/10.1088/1367-2630/aaecee} {\bibfield  {journal}
  {\bibinfo  {journal} {New J. Phys.}\ }\textbf {\bibinfo {volume} {20}},\
  \bibinfo {pages} {113024} (\bibinfo {year} {2018})}\BibitemShut {NoStop}%
\bibitem [{\citenamefont {Storn}\ and\ \citenamefont
  {Price}(1997)}]{StornJGlobalOptim1997}%
  \BibitemOpen
  \bibfield  {author} {\bibinfo {author} {\bibfnamefont {R.}~\bibnamefont
  {Storn}}\ and\ \bibinfo {author} {\bibfnamefont {K.}~\bibnamefont {Price}},\
  }\bibfield  {title} {\emph {\bibinfo {title} {Differential {{Evolution}}
  \textendash{} {{A Simple}} and {{Efficient Heuristic}} for global
  {{Optimization}} over {{Continuous Spaces}}}},\ }\href
  {https://doi.org/10.1023/A:1008202821328} {\bibfield  {journal} {\bibinfo
  {journal} {J. Global Optim.}\ }\textbf {\bibinfo {volume} {11}},\ \bibinfo
  {pages} {341} (\bibinfo {year} {1997})}\BibitemShut {NoStop}%
\bibitem [{\citenamefont {Virtanen}\ \emph {et~al.}(2020)\citenamefont
  {Virtanen}, \citenamefont {Gommers}, \citenamefont {Oliphant}, \citenamefont
  {Haberland}, \citenamefont {Quintero}, \citenamefont {Harris}, \citenamefont
  {Archibald}, \citenamefont {Ribeiro}, \citenamefont {Pedregosa},
  \citenamefont {{van Mulbregt}},\ and\ \citenamefont
  {{others}}}]{VirtanenNatMethods2020}%
  \BibitemOpen
  \bibfield  {author} {\bibinfo {author} {\bibfnamefont {P.}~\bibnamefont
  {Virtanen}}, \bibinfo {author} {\bibfnamefont {R.}~\bibnamefont {Gommers}},
  \bibinfo {author} {\bibfnamefont {T.~E.}\ \bibnamefont {Oliphant}}, \bibinfo
  {author} {\bibfnamefont {M.}~\bibnamefont {Haberland}}, \bibinfo {author}
  {\bibfnamefont {E.~A.}\ \bibnamefont {Quintero}}, \bibinfo {author}
  {\bibfnamefont {C.~R.}\ \bibnamefont {Harris}}, \bibinfo {author}
  {\bibfnamefont {A.~M.}\ \bibnamefont {Archibald}}, \bibinfo {author}
  {\bibfnamefont {A.~H.}\ \bibnamefont {Ribeiro}}, \bibinfo {author}
  {\bibfnamefont {F.}~\bibnamefont {Pedregosa}}, \bibinfo {author}
  {\bibfnamefont {P.}~\bibnamefont {{van Mulbregt}}},\ and\ \bibinfo {author}
  {\bibnamefont {{others}}},\ }\bibfield  {title} {\emph {\bibinfo {title}
  {{{SciPy}} 1.0: Fundamental algorithms for scientific computing in
  {{Python}}}},\ }\href {https://doi.org/10.1038/s41592-019-0686-2} {\bibfield
  {journal} {\bibinfo  {journal} {Nat Methods}\ }\textbf {\bibinfo {volume}
  {17}},\ \bibinfo {pages} {261} (\bibinfo {year} {2020})}\BibitemShut
  {NoStop}%
\bibitem [{\citenamefont {Kubala}\ \emph {et~al.}(2020)\citenamefont {Kubala},
  \citenamefont {Ankerhold},\ and\ \citenamefont
  {Armour}}]{KubalaNewJPhys2020}%
  \BibitemOpen
  \bibfield  {author} {\bibinfo {author} {\bibfnamefont {B.}~\bibnamefont
  {Kubala}}, \bibinfo {author} {\bibfnamefont {J.}~\bibnamefont {Ankerhold}},\
  and\ \bibinfo {author} {\bibfnamefont {A.~D.}\ \bibnamefont {Armour}},\
  }\bibfield  {title} {\emph {\bibinfo {title} {Electronic and photonic
  counting statistics as probes of non-equilibrium quantum dynamics}},\ }\href
  {https://doi.org/10.1088/1367-2630/ab6eb0} {\bibfield  {journal} {\bibinfo
  {journal} {New J. Phys.}\ }\textbf {\bibinfo {volume} {22}},\ \bibinfo
  {pages} {023010} (\bibinfo {year} {2020})}\BibitemShut {NoStop}%
\end{thebibliography}
\end{document}